\documentclass[lettersize,journal]{IEEEtran}
\usepackage{amsmath,amssymb,amsfonts}
\usepackage{array}
\usepackage[caption=false,font=normalsize,labelfont=sf,textfont=sf]{subfig}
\usepackage{textcomp}
\usepackage{stfloats}
\usepackage{url}
\usepackage{verbatim}
\usepackage{graphicx}
\usepackage{cite}
\usepackage{tabularray}
\usepackage{subfig}
\usepackage[table,xcdraw]{xcolor}
\usepackage{float} 
\usepackage{multirow}
\usepackage{textcomp}
\usepackage{xcolor}
\usepackage[table,xcdraw]{xcolor}
\usepackage{tabto}
\usepackage{tablefootnote}
\usepackage{breqn}
\usepackage{lipsum}
\usepackage{multirow}
\usepackage{colortbl}
\usepackage{hhline}
\usepackage{booktabs}
\usepackage[super]{nth}
\usepackage{nicematrix,enumitem}
\usepackage[para]{threeparttable}
\usepackage[noend]{algpseudocode}
\usepackage{algorithm}
\usepackage{makecell}

\begin{document}

\title{Margin Propagation based XOR-SAT Solvers for Decoding of LDPC Codes}

\author{Ankita Nandi,~\IEEEmembership{Student Member,~IEEE,} 
		Shantanu Chakrabartty,~\IEEEmembership{Senior Member,~IEEE}\\
		and~Chetan Singh Thakur,~\IEEEmembership{Senior Member,~IEEE}
}



\maketitle

\begin{abstract}
Decoding of Low-Density Parity Check (LDPC) codes can be viewed as a special case of XOR-SAT problems, for which low-computational complexity bit-flipping algorithms have been proposed in the literature. However, a performance gap exists between the bit-flipping LDPC decoding algorithms and the benchmark LDPC decoding algorithms, such as the Sum-Product Algorithm (SPA). In this paper, we propose an XOR-SAT solver using log-sum-exponential functions and demonstrate its advantages for LDPC decoding. This is then approximated using the Margin Propagation formulation to attain a low-complexity LDPC decoder. The proposed algorithm uses soft information to decide the bit-flips that maximize the number of parity check constraints satisfied over an optimization function. The proposed solver can achieve results that are within $0.1$dB of the Sum-Product Algorithm for the same number of code iterations. It is also at least $10 \times$ lesser than other Gradient-Descent Bit Flipping decoding algorithms, which are also bit-flipping algorithms based on optimization functions. The approximation using the Margin Propagation formulation does not require any multipliers, resulting in significantly lower computational complexity than other soft-decision Bit-Flipping LDPC decoders. 
\end{abstract}


\section{Introduction}
\begin{figure}[t]
	\centering
 	\subfloat[]
 {\includegraphics[width=0.3\linewidth]{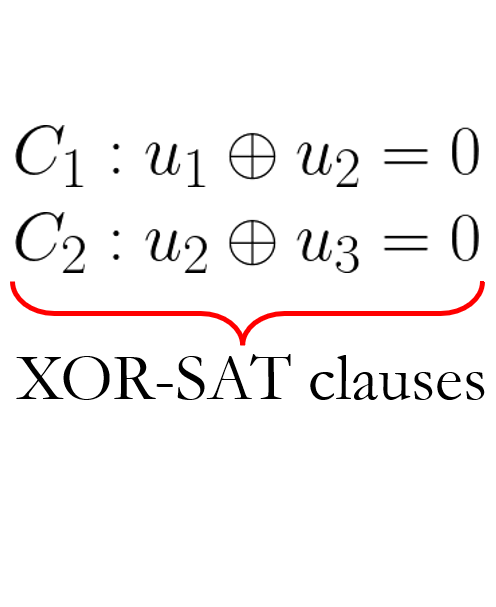}
		\label{concept_a}}
	\subfloat[]
   {\includegraphics[width=0.7\linewidth]{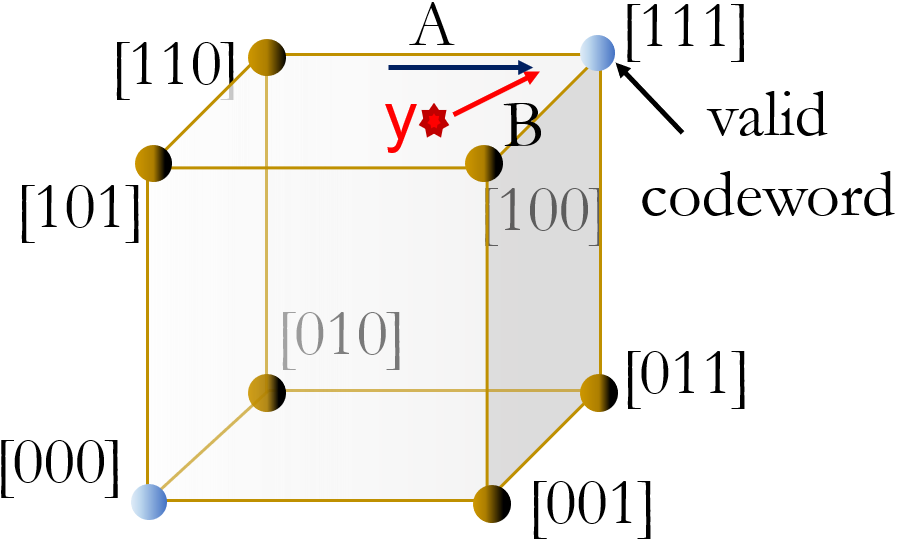}
		\label{concept_b}}

  	\subfloat[]
   {\includegraphics[width=0.99\linewidth]{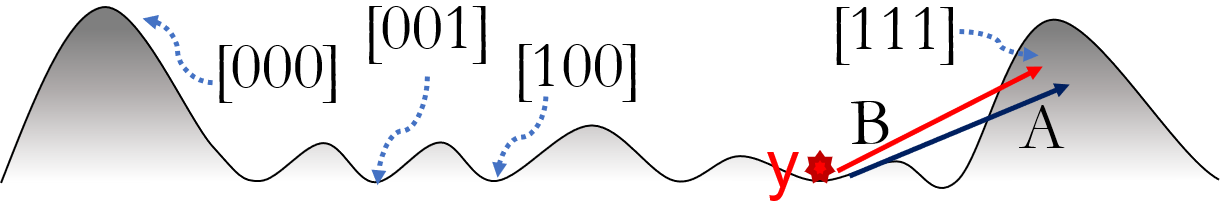}
		\label{concept_c}}
  
\caption{XOR-SAT as an equivalent optimization problem illustrated for a 3-bit majority code:\protect\subref{concept_a}~ The parity check constraints or XOR-clauses $C_1$ and $C_2$ corresponding to Boolean variables $u_1,u_2$ and $u_3$; \protect\subref{concept_b}~ Three-dimensional lattice showing the location of the assignments that satisfy $C_1$ and $C_2$ and hence represent valid codewords; \protect\subref{concept_c}~XOR-SAT or decoding as a continuous optimization problem where the global maxima correspond the solution or valid codewords.}
\label{concept}
\end{figure}

Maximum-Likelihood (ML) decoding of Low-Density Parity Check (LDPC) codes~\cite{gallager} and other linear block codes is generally considered to be an NP-hard problem~\cite{ldpc_nphard, ldpc_nphard2}~\cite{linear_nphard2003,vardy1997algorithmic}. As a result, practical decoders are implemented using sub-optimal methods. The NP-hardness of LDPC decoding arises because the number of Boolean variables involved in parity check conditions or XOR-clauses exceeds the number of clauses~\cite{book_shu_lin,molnar_maxsat}, leading to an under-determined set of linear equations. It may be noted that when the number of variables equals the number of clauses, the solution to the XOR satisfaction (XOR-SAT) problem is unique and can be determined in polynomial time. 

As an example, a $3-$variable XOR-SAT corresponding to the majority encoder is shown in Fig.~\ref{concept_a}. $C_1$ and $C_2$ represent the parity check constraints over the Boolean variables $u_1,u_2, u_3 \in \{0,1\}$. The solution to these XOR clauses can be depicted using a discrete lattice in Fig.~\ref{concept_b} where each of the lattice nodes represents a possible assignment to ($u_1, u_2, u_3$). An XOR-SAT solver produces an assignment that satisfies all the clauses $C_1$ and $C_2$, which in this case are valid codes [$000$] and [$111$].

LDPC decoders are a special case of XOR-SAT solvers where the objective is to reach a valid code that is closest (in some distance metric) to an initial (or received) vector (shown by $\mathbf{y}$ in Fig.~\ref{concept_b}). For a hard decoding approach, the starting vector is one of the lattice nodes, and a Hamming distance metric is used to evaluate the nearest valid codeword. For a soft decoding approach, the initial vector is real-valued and can lie within the volume of the lattice (denoted by $\mathbf{y}$ in Fig.~\ref{concept_b} and Fig.~\ref{concept_c}). Searching for the nearest valid codeword can be viewed as an optimization problem, as shown in Fig.~\ref{concept_c}, where the valid codewords are mapped as the maxima of an objective function and the invalid codewords as the minima~\cite{GDBF_wadayama}. Decoding then proceeds by an iterative procedure that progressively approaches the maximum that is nearest to the initial starting point. However, when the optimization landscape is non-convex, the number of local maxima scales exponentially with the number of variables. Therefore, optimal LDPC decoding is equivalent to solving a global optimization problem with an exponentially large search space.

Practical LDPC decoders resort to a sub-optimal iterative search algorithm, and in literature, the procedure can mainly be classified as a hard decision-based decoding algorithm, called the bit-flipping (BF) algorithm, or as soft decision-based Belief Propagation (BP) decoding, encompassing the Min-Sum (MS) and the Sum-Product Algorithm (SPA)~\cite{sarah_johnson_2009}. The key mechanism for these algorithms is the iterative passing of messages between two types of nodes, which are the variable nodes (corresponding to the bit nodes) and the check nodes (corresponding to the parity check clauses)~\cite{book_shu_lin}. The bit-flipping algorithms transmit binary message values to fulfill parity check equations and are characterized by their low complexity but limited performance. In contrast, the Belief Propagation-based algorithms, which take into account the reliability of bits, have demonstrated superior performance albeit at the cost of increased complexity~\cite{sarah_johnson_2009}. All other decoders can be considered as a combination of these two approaches.

\begin{figure*}[t]
	\centering
 {\includegraphics[width=0.99\linewidth]{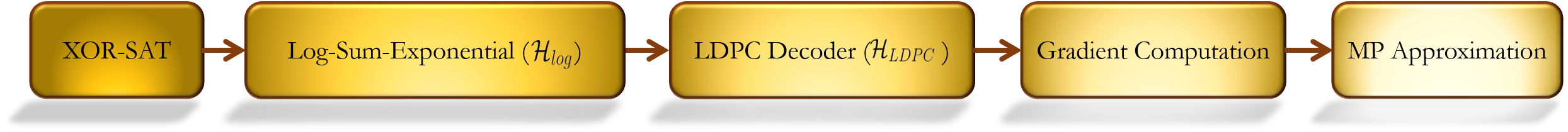}}

\caption{ Flow-chart showing key steps involved in the design of the MP-XOR-SAT based LDPC decoder.
}
\label{flow}
\end{figure*}

However, the optimization framework for decoding leads to a class of bit-flipping algorithms that exploit optimization landscape gradients to guide the iterative procedure. These are popularly called the Gradient Descent Bit Flipping (GDBF) algorithms. 
While most GDBF algorithms use hard decisions for a Binary Symmetric Channel (BSC), very few exploit the soft information that is available at the output of the communication channel. 

In this paper, we present a low-complexity LDPC decoder based on a Margin-Propagation (MP) based XOR-SAT solver that can achieve results within $0.1$dB of the Sum-Product Algorithm for an Additive White Gaussian  Noise (AWGN) channel. We call this the MP-XOR-SAT solver. The approach we take for designing the MP-XOR-SAT solver is illustrated in Fig.~\ref{flow}. The first step in the process is designing an objective function for XOR-SAT using the log-sum-exponential function, $\mathcal{H}_{log}$, which eliminates multiplications from the objective function. The second step is achieved by deriving $\mathcal{H}_{LDPC}$ for LDPC decoding. This is done by establishing the normalization factor alongside $\mathcal{H}_{log}$, which maintains a close correlation with the a-prior information (or channel information), thus aiding in achieving better decoding performance. 
The third step computes the gradient of the $\mathcal{H}_{LDPC}$ function and replaces the probability scores in the gradient with a Margin Propagation based probability scores. Additional heuristics are then incorporated in the Margin Propagation based formulation, which leads to the proposed MP-XOR-SAT based LDPC decoder. 
Based on this design flow, the proposed work has the following key contributions:

\begin{itemize}
    \item The proposed framework, called MP-XOR-SAT, uses a piece-wise linear approximation called Margin Propagation to implement XOR-SAT solvers.
    
    \item The resulting MP-XOR-SAT decoders do not require multiplications, leading to a much lower computational and hardware implementation complexity.

    \item MP-based XORSAT (MP-XOR-SAT) algorithm, when used for LDPC decoding, takes $10 \times$ fewer code iterations than state-of-the-art Gradient Descent Bit Flipping algorithms and the same code iterations as that of Sum-Product Algorithm (SPA), with the error rates being within $0.1$dB of Sum-Product Algorithm (SPA).
 
\end{itemize}
The paper is organized as follows. A literature survey of the state-of-the-art Gradient Descent Bit Flipping algorithms and prior work on Margin Propagation algorithms for LDPC decoding is presented in Section~\ref{related}. Section~\ref{LSE_xorsat} defined the XOR-SAT problem using log-sum-exponential functions, and in Section~\ref{ldpc_decoder}, the XOR-SAT problem is tailored to suit LDPC decoding followed by its MP approximation. The results and discussion are in Section~\ref{results} and Section~\ref{discussion}, respectively. Finally, the paper is concluded in Section~\ref{conclusion}.

\section{Related Works}\label{related}
It is worth noting that both Gradient Descent Bit Flipping (GDBF) algorithms and Margin Propagation (MP) have been previously used to implement LDPC decoders. Thus, the contributions and challenges faced by these techniques have been discussed in this section. 

\subsection{Gradient Descent Bit Flipping (GDBF) for LDPC codes}
 
Bit-flipping algorithms as an optimization problem have first been considered in~\cite{GDBF_wadayama}, where Gradient Descent Bit Flipping (GDBF) was proposed. Their objective function included two terms, one being the constraint satisfaction and the other to check for the word bearing maximum correlation with the Maximum Likelihood decoding. Both the single-bit and multi-bit flipping mechanisms were proposed, including an escape mechanism to release the function from local optima points. There were various spin-offs~\cite{multibitGDBF,improvedGDBF} that caused the Gradient Descent Bit Flipping algorithm to become a separate class of decoding algorithms. However, the results showed that the Min-Sum algorithm performed better than the proposed Gradient Descent Bit Flipping for both single and multi-bit flip versions. 

In~\cite{ngdbf}, a low-complexity solution is proposed by introducing random perturbations. The objective function not only includes terms for the constraint satisfaction check and maximum correlation to the received but also adds random noise. However, these random perturbations are introduced based on the noise in the channel  (with variance $~N_o/2$). Both single-bit and multi-bit variations were demonstrated on the Additive White Gaussian Noise channel. This algorithm comes to $0.5$dB of the Belief Propagation algorithm. 

The Probabilistic Gradient Descent Bit Flipping (PGDBF)~\cite{pgdbf} models the problem to a Binary Symmetric Channel (BSC). In this decoder, a randomly chosen fraction of bits is flipped, leading to a large performance improvement. The authors in~\cite{pgdbf_flash} present a Flash Memory-adapted PGDBF decoder and show the results of each of them using BSC. A variant of PGDBF, which keeps track of the trapping sets for the flipping, was proposed in~\cite{t-pgdbf}. The Information Storage Bit Flipping (ISBF)~\cite{isbf} is mainly focused on optimizing the critical path in finding the global maxima in order to improve throughput. 

The Gradient Descent Bit Flipping algorithms with Momentum (GDBFwM)~\cite{gdbf_wm} memorizes the update of the bits at every iteration, thus providing inertia to the direction of the update. The results depict a striking resemblance in performance with respect to the Belief Propagation. The AD-GDBF decoders proposed in~\cite{ad-gdbf} show that for LDPC codes and a Binary Symmetric Channel, the frame error rates are better than the Belief Propagation without the knowledge of trapping sets. A very recent syndrome bit flipping (SBF) algorithm is proposed~\cite{gdbf_2023}, which only takes into account the syndrome and is used together with the NGDBF algorithm to yield results almost $0.5$dB better than belief propagation but with $300$ iterations for NGDBF and $100$ iterations for SBF.

However, most of the earlier works have a similar form of objective function and can be generalized as a single function, which will be shown in Section~\ref{discussion}. In this work, we propose an objective function that is not equivalent to any of the other represented functions. Further, the candidate objective function will be approximated using the Margin Propagation algorithm to reduce computational and hardware complexity. A brief overview of the Margin Propagation algorithm and its LDPC-related works is discussed in the next section.  

\subsection{Margin Propagation based LDPC decoders}

The Margin Propagation (MP) principle was first introduced as a mechanism to approximate different variants of log-sum-exponential functions using only piecewise-linear functions~\cite{chakrabartty_thesis}. The resulting circuits and systems obviated the use of multipliers and could be easily mapped onto different hardware substrates~\cite{minggu_thesis}. The approach has since been used to implement analog classifiers~\cite{kucher2007energy,synth}, LDPC decoders~\cite{gu2011adaptive}, digital accelerators~\cite{abhishek}, analog correlators~\cite{karimisscc} and field-programmable machine learning processors~\cite{pk1,pk2}. 

Gu. et al. in~\cite{ldpc_minggu} have demonstrated a Margin Propagation based $(32,8)$ regular LDPC code. The formulation was essentially differential, which became more complex as the sizes of the codeword and the degrees of the check nodes and the variable nodes increased. 
A probabilistic XOR-based check node implementation that uses the Generalized Margin Propagation~\cite{pk1} circuit was proposed for the Sum-Product Algorithm in~\cite{vlsid}, but it requires the translation of the Sum-Product Algorithm into the probabilistic domain. 

In this work, we first propose a log-sum-exponential based XOR-SAT solver, which is then used for LDPC decoding.


\begin{table}[t]
\centering
\caption{Notations}
\scalebox{0.99}{
\begin{tabular}{|l|l|}
\hline
Symbol & Notation   \\ \hline
$A$ & Scalar   \\
$\mathbf{a}$ & Vector   \\
$\mathbf{A}$ & Matrix   \\
$a_{i,j}$ &  $(i,j)^{th}$ element of the matrix $\mathbf{A}$  \\
$a_j$ & $j^{th}$ Element in a vector $\mathbf{a}$   \\
$|A|$ & Absolute value of the scalar $A$   \\
$\mathcal{P}$ & Set    \\
$|\mathcal{P}|$ & Cardinality of the set $\mathcal{P}$   \\
$\{\phi\}$ & Empty set with $|\{\phi\}| = 0$   \\
$\mathbb{I} (b)$ & Indicator function for the Boolean condition b  \\

\hline
\end{tabular}}
\label{notations}
\end{table}

\section{XOR-SAT Optimization}\label{LSE_xorsat}
We first formulate XOR-SAT as an optimization problem using log-sum-exponential functions. 
Our mathematical exposition will use the notations summarized in Table~\ref{notations}, and all the operators will be defined as they are introduced. Since our goal is to use the MP-XOR-SAT solver for LDPC decoding, we will conform to the notations used in the LDPC literature to describe the XOR-SAT clauses and constraints. 

Let $N$ be the number of Boolean variables $u_1, u_2, \dots, u_{N} \in \mathbb{F}_2$ and $M$ be the number of XOR clauses or parity check clauses.
The XOR clauses $\mathbf{C} = \{C_1, C_2 \cdots, C_M\}$ can be compactly expressed using a parity check matrix $\mathbf{H} \in \mathbb{F}^M_2 \times \mathbb{F}^N_2$ as $\mathbf{C} = \mathbf{Hu}$ with $\mathbf{u} = (u_1, u_2, \cdots, u_{N})$ being a vector comprising of all the Boolean variables. 
Here, the clause $C_i$ is $C_i~=~\underset{j:h_{ij} = 1}{\oplus} u_j$, which denotes an XOR combination of all participating Boolean variables. The $(i,j)^{th}$ element of $\mathbf{H}$, i.e., $h_{i,j} = 1 (0)$ indicates if $u_j$ is participating (not participating) in the clause $C_i$.  We denote $\mathcal{P}^+$ as the set of clauses $C_i$ that satisfy equation~\eqref{parity_eq_sat} and $\mathcal{P}^-$ as the set of clauses $C_i$ that are not satisfied (i.e., equation~\eqref{parity_eq_no_sat}). The XOR-SAT problem is to identify an assignment $\mathbf{u}$ that satisfies $\mathbf{C}=0$. 

A general approach to solving XOR-SAT problems is using a bit-flipping procedure, which is summarized by the pseudo-code in Algorithm~\ref{pseudocode}. The algorithm iteratively identifies the variable that influences most of the unsatisfied clauses and then flips the state of that variable. The uniqueness of a bit-flipping algorithm is the procedure to identify the variables to flip. The proposed design of MP-XOR-SAT is one such procedure, which will be described next.

We introduce a one-to-one bipolar mapping $d_i = (2u_i-1)$ such that $d_i \in \{-1,+1\}$. Therefore, when the constraint $C_i$ is satisfied or $C_i \in \mathcal{P}^+$, it can be written as, 
\begin{equation}
   \underset{j:h_{ij} = 1}{\oplus} u_j =  0  \implies \prod_{j=1}^{N}d_j^{h_{i,j}}  = 1  
   \label{parity_eq_sat}
\end{equation}
whereas when $C_i$ is not satisfied, $C_i \in \mathcal{P}^-$, and can be written as
\begin{equation}
   \underset{j:h_{ij} = 1}{\oplus} u_j =  1  \implies \prod_{j=1}^{N}d_j^{h_{i,j}}  = -1. 
   \label{parity_eq_no_sat}
\end{equation}

\begin{algorithm}[t]
\caption{General Bit Flipping Approach}
\label{pseudocode}
\begin{algorithmic}[1]
\State \textbf{Constraint Clauses} \\
$C_1$: $\underset{j:h_{1j} = 1}{\oplus} u_j = 0$\\
$C_2$: $\underset{j:h_{2j} = 1}{\oplus} u_j = 0$\\
$\vdots$\\
$C_M$: $\underset{j:h_{Mj} = 1}{\oplus} u_j = 0$\\

\State \textbf{Conditions} \\
$\mathcal{P}^+$: Set of satisfied clauses\\
$\mathcal{P}^-$: Set of unsatisfied clauses\\

\State \textbf{Pseudo Code}

\State Find the variable $u_j$ that most influences $\mathcal{P}^-$
\State Flip the state of $u_j$
\State Continue until all clauses are in $\mathcal{P}^+$ or maximum iterations ($I_{max}$) have reached
\State

\State \textbf{Output:} Decision when $|\mathcal{P}^-| = 0$ or $I = {(I_{max})}$ where $I$ refers to the current iteration 
\end{algorithmic}
\end{algorithm}


Therefore, all XOR-clauses are satisfied if and only if  
\begin{equation}
\sum\limits_{i=1}^{M}  \prod\limits_{j=1}^{N}d_j ^ {h_{i,j}} = M.
\label{valid1}
\end{equation}
We now introduce auxiliary variables $x_j \in [0,1]$, $\forall j = 1,..,N$ such that the expression
\begin{equation}
\max_{x_j \in [0,1]} \sum\limits_{i=1}^{M} \left( \prod\limits_{j=1}^{N}d_j ^ {h_{i,j}} \prod_{j=1}^{N}x_j ^ {h_{i,j}}\right) = M.
\label{valid2}
\end{equation}
is also satisfied if and only if all XOR-clauses are satisfied. Equation~\eqref{valid2} can also be written as
\begin{equation}
\max_{x_j \in [0,1]}  \sum\limits_{i:C_i \in \mathcal{P}^+} \left ( \prod\limits_{j = 1}^{N}x_j^{h_{i,j}}\right ) = M.
\label{valid3}
\end{equation}

We can frame equation~\eqref{valid3} as a solution to the following combinatorial optimization problem that also includes the set of
unsatisfied clauses $\mathcal{P}^-$:

\begin{equation}
    \max_{\mathcal{P}^+, \mathcal{P}^-, x_j \in [0,1]} \frac{\sum\limits_{i:C_i \in \mathcal{P}^+} \left ( \prod\limits_{j = 1}^{N}x_j^{h_{i,j}}\right )}{\sum\limits_{i:C_i \in \mathcal{P}^-} \left ( \prod\limits_{j = 1}^{N}x_j^{h_{i,j}}\right )}
    \label{check_loss}
\end{equation}
Here, the optimization is over the discrete sets $\mathcal{P}^+,\mathcal{P}^-$ implies optimization over the discrete variables $d_j$. Note that the optimization in equation~\eqref{check_loss} achieves its maximum only when all the clauses are satisfied or $\mathcal{P}^- = \{\phi\}$. 

Using the monotonic property of the $\log(.)$ function, the optimization in equation~\eqref{check_loss} is equivalent to
 
\begin{eqnarray}
\label{check_loss2}
    \max_{\mathcal{P}^+, \mathcal{P}^-, x_j \in [0,1]}  \log \sum\limits_{i:C_i \in \mathcal{P}^+} \prod_{j = 1}^{N}x_j^{h_{i,j}} \ldots\\ \nonumber
    -  \log  \sum\limits_{i:C_i \in \mathcal{P}^-} \prod_{j = 1}^{N}x_j^{h_{i,j}} 
\end{eqnarray}

Expressing equation~\eqref{check_loss2} using $q_j = \log x_j$ in equation~\eqref{check_loss3}, we have
\begin{eqnarray}
\label{check_loss3}
         \max_{\mathcal{P}^+, \mathcal{P}^-, q_j \in [-\infty,0]}   \log \left[\sum\limits_{i:C_i \in \mathcal{P}^+} \exp\left(\sum\limits_{j = 1}^{N} h_{i,j} q_j \right) \right] \ldots\\ \nonumber
    -  \log \left[ \sum\limits_{i:C_i \in \mathcal{P}^-}\exp\left(\sum\limits_{j = 1}^{N} h_{i,j} q_j \right) \right]
\end{eqnarray}

\algblock{ForMod}{EndForMod}
\algnewcommand\algorithmicparfor{\textbf{for}}
\algnewcommand\algorithmicpardo{\textbf{}}
\algnewcommand\algorithmicendparfor{\textbf{end-for}}
\algrenewtext{ForMod}[1]{\algorithmicparfor\ #1\ \algorithmicpardo}
\algrenewtext{EndForMod}{\algorithmicendparfor}
\algblock{WhileMod}{EndWhileMod}
\algnewcommand\algorithmicwhilenew{\textbf{while}}
\algnewcommand\algorithmicdonew{\textbf{}}
\algnewcommand\algorithmicendwhilenew{\textbf{end-while}}
\algrenewtext{WhileMod}[1]{\algorithmicwhilenew\ #1\ \algorithmicdonew}
\algrenewtext{EndWhileMod}{\algorithmicendwhilenew}
\begin{algorithm}[htbp]
\caption{MP-XOR-SAT LDPC Decoding Algorithm }
\label{shapemapping}
\begin{algorithmic}[1]
\State \textbf{Inputs:} \\
Channel Output: $ r_j$ $\in \mathbb{R}$, $\forall j = 1, 2, \cdots, N$\\
Parity Check Matrix: $H = \left[h_{i,j}\right] \forall i=1,..,M, j=1,..,N$\\
Maximum Iterations: $I_{max}$\\

\State \textbf{Output:} \\
Decision: $\mathbf{u} \in \{0, 1\}^N$\\

\State \textbf{Hyper-parameters:} \\
$\tau$: MP constraint constant\\
$\theta$: Decision Threshold\\
$\eta$: Learning rate\\
\State \textbf{Initialization:}\\
$d_j \leftarrow sign(r_j)$ \\
$q_j \leftarrow d_j \log\left(|tanh(r_j)|\right)$ \\

\State \textbf{Algorithm:}
\WhileMod{(I $ \leq I_{max}$)}
    
    \State $\mathcal{P}^+ = \{C_i: \prod\limits_{j=1}^{N}d_j ^ {h_{i,j}} = 1\}$\textit{~~~\#Set of satisfied clause}
    \State $\mathcal{P}^- = \{C_i: \prod\limits_{j=1}^{N}d_j ^ {h_{i,j}} = -1\} $\textit{~\#Set of unsatisfied clause}
    \State 
  \State \textbf{if} $\left (|\mathcal{P}^-| = 0 \right )$ \textit{~~~~~~~~~~~~~~~~~~~~~~~~\#Terminate iterations}
                    \State \hspace{1.5em} \textbf{break};
   \State \textbf{end}
  \State 
    \ForMod{$i =  1:M$}
    \State $z_i = \sum\limits_{j =1}^N h_{i,j}q_j$
            \State \textbf{if} $\left (C_i \in \mathcal{P}^+ \right ) $ \textit{~~~~~~~~~~~~~~~~~\#Determine $C_i \in \mathcal{P}^+$}
                    \State \hspace{3em}$z_i^+ = z_i$
                    \State \hspace{3em}$z_i^- = q_{min}$
            \State \textbf{else}\textit{~~~~~~~~~~~~~~~~~~~~~~~~~~~~~~\#Determine $C_i \in \mathcal{P}^-$}
                    \State \hspace{3em}$z_i^- = z_i$
                    \State \hspace{3em}$z_i^+ = q_{min}$
            \State \textbf{end-if}
  \EndForMod\\

  \State $\zeta^+ = MP(\mathbf{z}^+,\tau)$
  \State $\zeta^- = MP(\mathbf{z}^-,\tau)$

  \ForMod{$j =  1:N$}

    \State \textbf{if} $q_j < \theta$, 
     \State \hspace{1.5em} $d_j \leftarrow -d_j$\textit{~~~~~~~~~~~~~~~~~~~~~~~~~~~~~~~~~~~~\#Bit flip}
     \State \textbf{end-if}
     \State
     \State \textit{\#Gradient Update}
     \State $q_j \leftarrow q_j + \eta \left( \sum\limits_{i=1}^{M} h_{ij}\frac{ [z_i^+-\zeta^+]_+- [z_i^--\zeta^-]_+}{\tau \times \mathbb{A}_j} + r_jd_j\right)$

  \EndForMod\\
  \State I = I+1
\EndWhileMod

\State \textbf{Decoded Output:} $u_j = d_j > 0$ $\forall j = 1,2, \cdots, N$ 
\end{algorithmic}
\end{algorithm}

Introducing a small number $\epsilon > 0$ to ensure numerical stability for the conditions when either $|\mathcal{P}^+| = 0$, or $|\mathcal{P}^-| = 0 $, equation~\eqref{check_loss3} is written as 

\begin{eqnarray}
    \max_{\mathcal{P}^+, \mathcal{P}^-, q_j \in [-\infty,0]} \mathcal{H}_{log}
    \label{check_loss4}
\end{eqnarray}
where

\begin{eqnarray}
\label{hlog}
      \mathcal{H}_{log} =   \log \left[\sum\limits_{i:C_i \in \mathcal{P}^+}  \exp\left(\sum\limits_{j = 1}^{N} h_{i,j} q_j \right) + \epsilon  \right]  \ldots\\ \nonumber 
    -   \log \left[ \sum\limits_{i:C_i \in \mathcal{P}^-}  \exp\left(\sum\limits_{j = 1}^{N} h_{i,j} q_j \right) + \epsilon  \right]  
\end{eqnarray}

This is one candidate function for solving XOR-SAT problems.
We will now use this function denoted by $\mathcal{H}_{log}$ in equation~\eqref{hlog} to design LDPC decoders.

\section{MP-XOR-SAT based LDPC Decoder}\label{ldpc_decoder}

To apply the formulation in $\mathcal{H}_{log}$ for decoding LDPC codes, it is imperative to not only receive a solution that satisfies all the clauses but to ensure that the decoded codeword matches the transmitted codeword. The number of code bits in the codeword is the number of XOR-SAT Boolean variables $N$, and $M$ is the number of XOR clauses described by the parity-check matrix $\mathbf{H}$. In LDPC parlance, the parity check matrix $\mathbf{H}$ consists of $M$ check nodes and $N$ variable nodes. 

Let us denote $\mathbf{c} = (c_1, c_2, \dots, c_{N})$ to be the transmitted codeword such that $\{ c_i \in F^N_2: \mathbf{Hc}=0\}$. By convention, the codeword $\mathbf{c}$ is mapped onto a BPSK (Binary Phase Shift Keying) signal constellation before transmission such that $\mathbf{\hat{c}} = (\hat{c}_1, \hat{c}_2, \dots, \hat{c}_{N}), \hat{c}_i = (2c_i-1)$. If the symbols $\mathbf{\hat{c}}$ are transmitted over an Additive White Gaussian Noise (AWGN) channel then $\mathbf{r} = (r_1, r_2, \dots, r_{N}), r_i \in \mathbb{R}$ is the channel output vector such that $\mathbf{r} = \hat{\mathbf{c}}  +\mathbf{n}$, where $\mathbf{n} \in \mathbb{R}^N$ is a random noise vector drawn from a Gaussian distribution~\cite{book_shu_lin}.

The information that is received from the channel can now be used to guide the dynamics of the MP-XOR-SAT solver such that it seeks a valid codeword that is close to the received channel vector $\mathbf{r}$.

To ensure this, we augment the $\mathcal{H}_{log}$ function in equation~\eqref{check_loss5} to include a term $\sum\limits_{j=1}^{N} r_j d_{j} q_j$ that captures the correlation between the decoded bits and the channel information. We call this the \textit{normalization factor} throughout the text. The procedure is similar to the maximum correlation decoding rule described in~\cite{GDBF_wadayama, ngdbf, gdbf_wm}. The augmented optimization function $\mathcal{H}_{LDPC}$ is given by:
\begin{equation}
    \mathcal{H}_{LDPC} = \mathcal{H}_{log} +\sum\limits_{j=1}^{N}r_j d_{j} q_j
    \label{check_loss5}
\end{equation}

The importance of the extrinsic correlation term in $\mathcal{H}_{LDPC}$ is to break the symmetry between the codewords such that the global maximum of $\mathcal{H}_{LDPC}$ corresponds to the codeword closest to the channel output. Thus, different types of LDPC decoders can be designed by maximizing $\mathcal{H}_{LDPC}$ according to: 
\begin{eqnarray}
    \max_{\mathcal{P}^+, \mathcal{P}^-, q_j \in [-\infty,0]} \mathcal{H}_{LDPC}
    \label{check_ldpc}
\end{eqnarray}
Next, we use a gradient ascent bit flipping approach to solve equation~\eqref{check_ldpc} and to design a decoder.

\begin{figure*}[t]
	\centering

  	\subfloat[]
 {\includegraphics[width=0.24\linewidth]{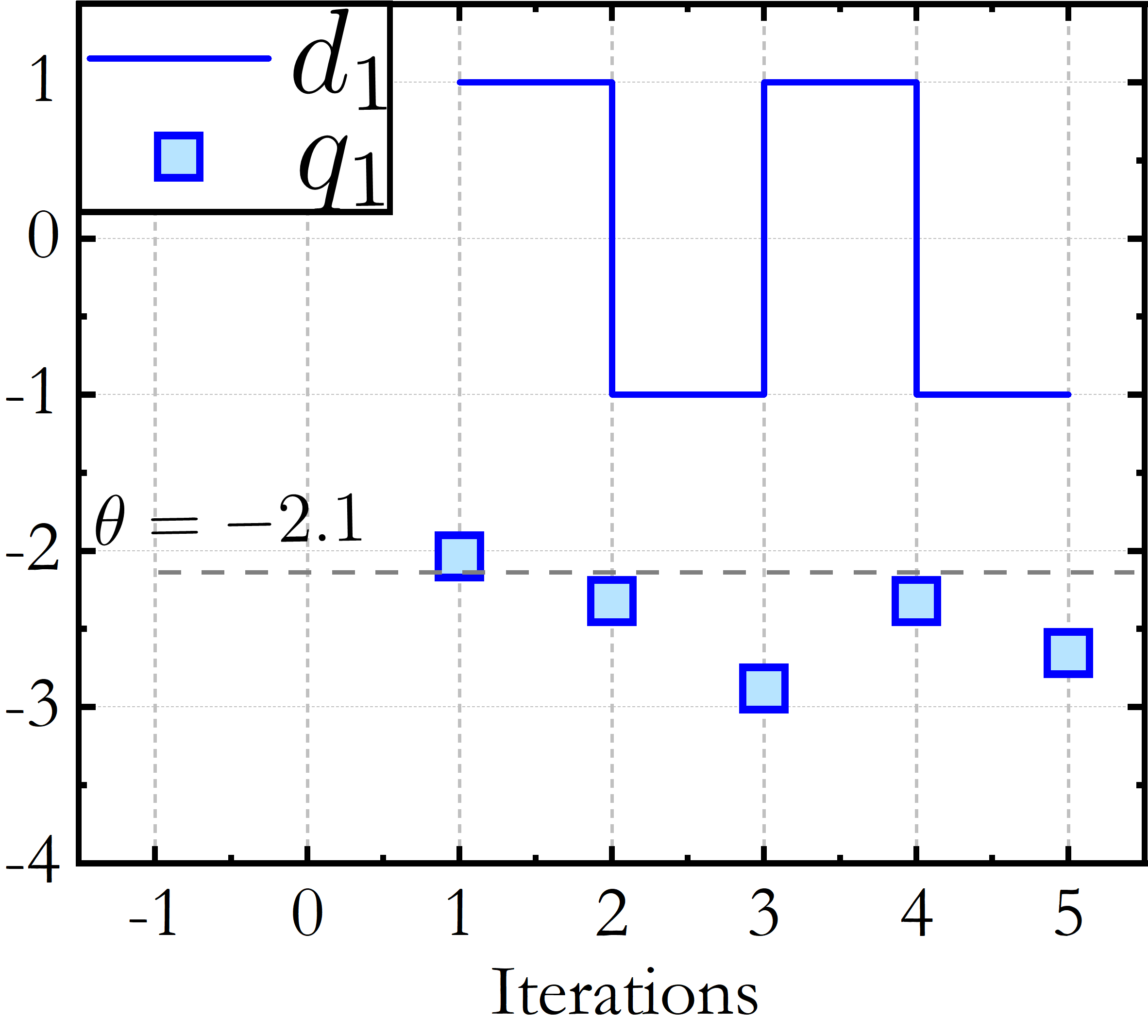}
		\label{eg_3bit_j1}}
	\subfloat[]
   {\includegraphics[width=0.24\linewidth]{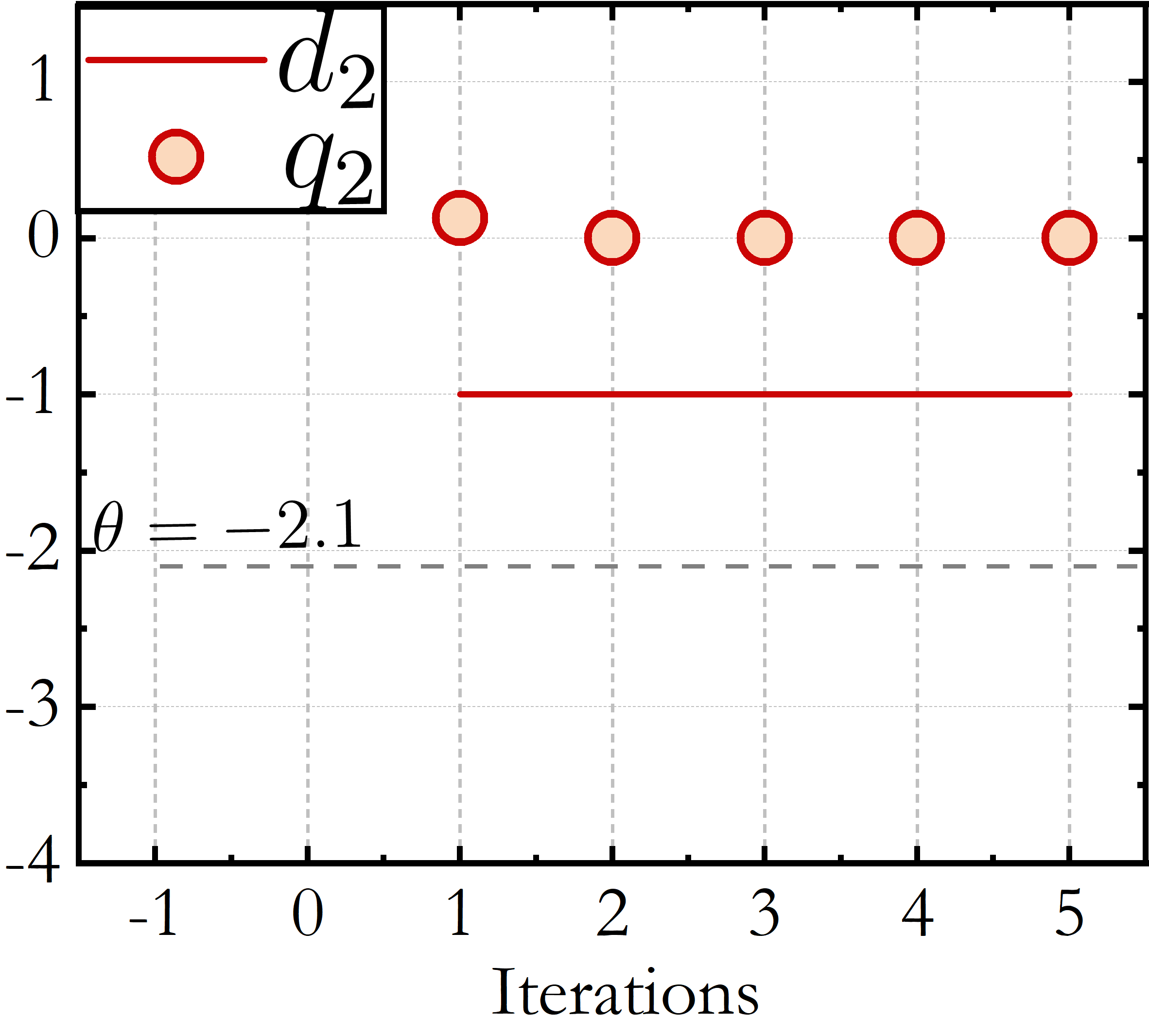}
		\label{eg_3bit_j2}}
  	\subfloat[]
   {\includegraphics[width=0.24\linewidth]{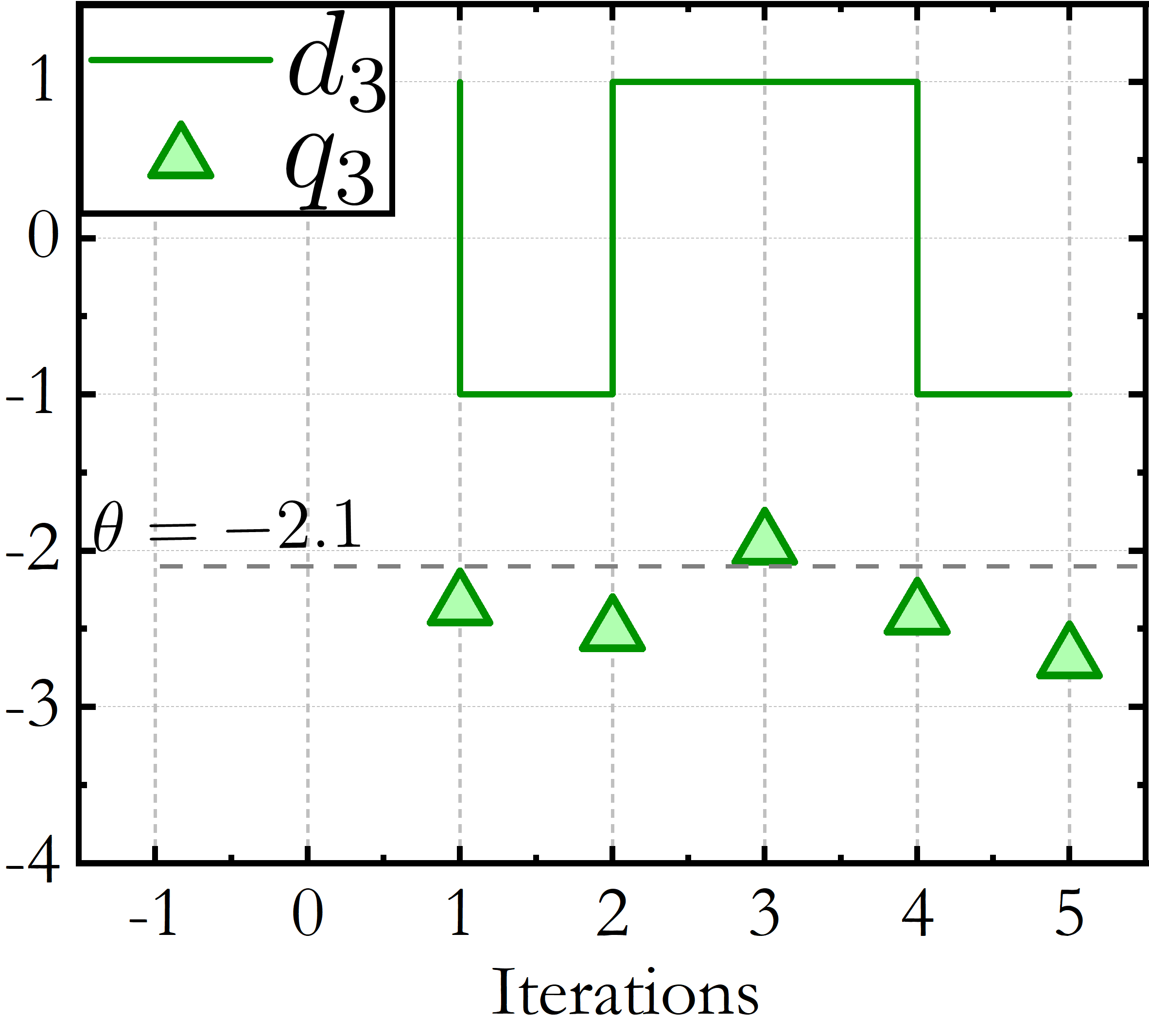}
		\label{eg_3bit_j3}}  	
  \subfloat[]
   {\includegraphics[width=0.26\linewidth]{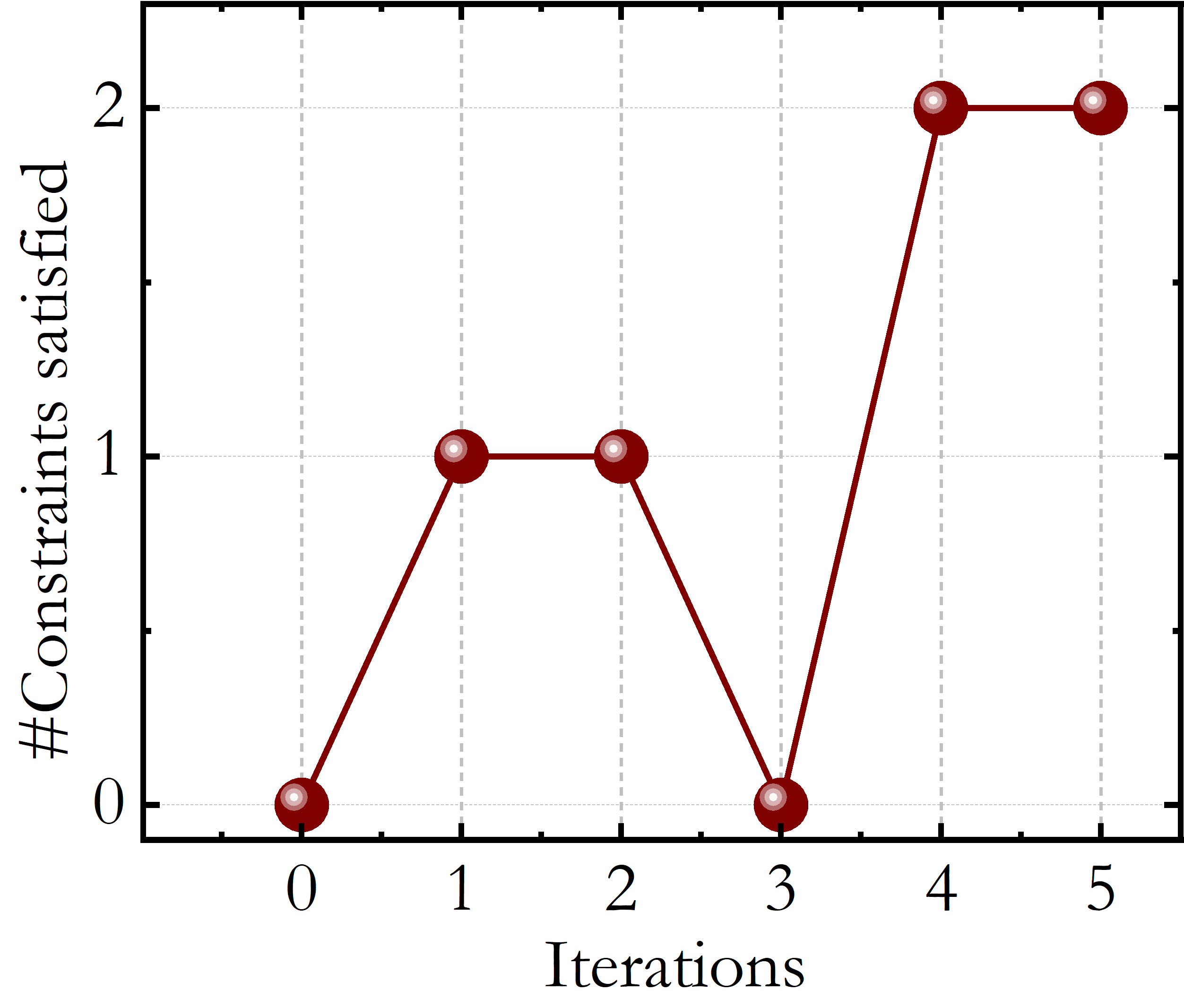}
		\label{eg_3bit_constraint}}

\caption{An example is illustrated using the 3-bit majority code. The evolution of the bits is explained over $5$ iterations, with $d_j$ being the decision bit and $q_j$ being the soft information. \protect\subref{eg_3bit_j1}~Dynamics of $d_1$, corresponding to the changes in $q_1$; \protect\subref{eg_3bit_j2}~Dynamics of $d_2$ due to the changes in $q_2$, which do not change because they do not satisfy the threshold condition; \protect\subref{eg_3bit_j3}~Dynamics of $d_3$ corresponding to the updates in $q_3$. \protect\subref{eg_3bit_constraint}~The number of constraints satisfied over each iteration. The $0^{th}$ iteration refers to the initial condition before the iterative algorithm is initiated.}
\label{eg_3bit}
\end{figure*}



\subsection{Gradient-ascent Updates and Bit-flips}
It may be noted that the function $\mathcal{H}_{LDPC}$ comprises both continuous variables $\mathbf{q}$ and discrete variables $\mathbf{d}$ that determine the membership of the sets $\mathcal{P}^+,\mathcal{P}^-$. Therefore, the gradient of $\mathcal{H}_{LDPC}$ is only defined when the memberships of the sets $\mathcal{P}^+,\mathcal{P}^-$ do not change. Hence, the proposed decoding algorithm will proceed as follows: 

At every iteration, the clauses are categorized into the sets $\mathcal{P}^+$ and $\mathcal{P}^-$ based on the state of the discrete variables $\mathbf{d}$. The continuous variables $\mathbf{q}$ are updated at each iteration according to a gradient ascent update rule 
\begin{eqnarray}
    \Delta \mathbf{q} = \eta \left. \nabla_{\mathbf{q}} \mathcal{H}_{LDPC} \right\vert_{\mathcal{P}^+,\mathcal{P}^- = const}
    \label{check_grad}
\end{eqnarray}
where $\eta > 0$ is the learning-rate parameter while the membership of the sets $\mathcal{P}^+$ and $\mathcal{P}^-$ remain unchanged.

If $z_i = \sum\limits_{i=1}^{N} h_{ij}q_j$, then based on the update equation~\eqref{check_grad} and equation~\eqref{check_loss5}, the gradient can be written as
\begin{eqnarray}
\label{diff_hldpc1_main}
\frac{\partial \mathcal{H}_{LDPC}}{\partial q_j} = \sum\limits_{i:C_i \in \mathcal{P}^+} h_{i,j}\left[\frac{\exp(z_i)}{\sum\limits_{i:C_i \in \mathcal{P}^+}\exp(z_i) + \exp(q_{min})} \right] \dots \\ \nonumber
- \sum\limits_{i:C_i \in \mathcal{P}^-} h_{i,j}\left[\frac{\exp(z_i)}{\sum\limits_{i:C_i \in \mathcal{P}^-}\exp(z_i) + \exp(q_{min})} \right] + r_jd_j
\end{eqnarray}
where $q_{min} = \log(\epsilon)$.
 Note that the multinomial logistic function $\frac{\exp(y_i)}{\sum\limits_{i}\exp(y_i)}$ in equation~\eqref{diff_hldpc1_main}, represents a probability score that arises due to the log-sum-exponential logistic cost function in equation~\eqref{hlog}. In our previous work~\cite{chakrabartty2007gini}, we have shown that other forms of cost functions and associated probability scores could be used as a replacement for the logistic cost function. These probability scores exhibit attractive properties, which include sparsity, faster convergence, and robustness~\cite{tnnlsoindrila}. Here, we will replace the logistic function by the Margin Propagation based probability score, which was introduced in~\cite{chakrabartty2007gini}. 
 
 Given an input vector $\mathbf{y} = \{y_i\}, i = 1,..,M$, the MP based probability score is given by
 \begin{equation}
    P_i = \frac{1}{\tau} [y_{i} - \zeta]_+
\end{equation}
where $[.]_+$ is a ReLU function, $\tau > 0$ is a hyper-parameter and $\zeta = MP(\mathbf{y},\tau)$ is an MP function that is computed to satisfy the normalization criterion
\begin{equation}
    \sum_{i = 1}^M P_i = 1
    \label{MP}
\end{equation}
The key advantage of using the Margin Propagation function and Margin Propagation probability scores is that it uses simple operations like ReLU and subtractions (rather than exponentiations and multiplications), and equation~\eqref{MP} can be easily implemented using analog and digital circuits~\cite{synth}.

To apply the Margin Propagation based probability scores, we define $z_i^+=z_i\cdot \mathbb{I}(C_i \in \mathcal{P}^+)$ and $z_i^-=z_i \cdot \mathbb{I}(C_i \in \mathcal{P}^-)$ for form vectors $\mathbf{z}^+,\mathbf{z}^-$. 

We then estimate the MP functions $\zeta^+ = MP(\mathbf{z}^+,\tau)$ and $\zeta^- = MP(\mathbf{z}^-,\tau)$. These approximations are used to replace the probability scores in equation~\eqref{diff_hldpc1_main}, which leads to
\begin{equation}
\frac{\partial \mathcal{H}_{LDPC}}{\partial q_j} =  \sum\limits_{i=1}^{M} h_{ij}\frac{[z_i^+-\zeta^+]_+-[z_i^--\zeta^-]_+}{\tau}  + r_jd_j.
    \label{diff_hldpc2_prev}
\end{equation}
The dynamics of the algorithm can be improved by introducing a scaling term $\mathbb{A}_j > 0, j = 1,..,N$, which controls the influence of the extrinsic factor $r_jd_j$ to the gradient update. Incorporating this heuristic, equation~\eqref{diff_hldpc2_prev} becomes 
\begin{equation}
\frac{\partial \mathcal{H}_{LDPC}}{\partial q_j} =  \sum\limits_{i=1}^{M} h_{ij}\frac{[z_i^+-\zeta^+]_+-[z_i^--\zeta^-]_+}{\tau \times \mathbb{A}_j}  + r_jd_j
    \label{diff_hldpc2_main}
\end{equation}

As a heuristic we chose the value of $\mathbb{A}_j$ as, $\mathbb{A}_j~=~\sum\limits_{i=1}^M h_{i,j}\left[\mathbb{I}(z_i^+ > \zeta^+) + \mathbb{I}(z_i^- > \zeta^-)\right]$ which measures the number of instances the variable $q_j$ is significant amongst clauses that appear in both the sets $\mathcal{P}^+,\mathcal{P}^-$. Thus, $\mathbb{A}_j$ being large implies the ambiguity in the influence $q_j$, in which case the gradient weighs the extrinsic information more. In the extreme case, when $\mathbb{A}_j = 1$, the variable $q_j$ appears significant only for one clause, in which case the gradient weighs the extrinsic information less. 

The change in $\Delta q_j = \eta \frac{\partial \mathcal{H}_{LDPC}}{\partial q_j}$ is then used to determine how to change the membership of the sets $\mathcal{P}^+,\mathcal{P}^-$ in the next iteration. For a given iteration, if the magnitude of any variable $q_j$ becomes lower than a threshold $\theta$, then the sign of the binary variable $d_j$ is flipped. The procedure of gradient updates and bit-flipping is repeated iteratively till all clauses are satisfied (i.e., the set $\mathcal{P}^-$ is empty) or the maximum allowable iterations ($I_{max}$) is reached. At the end of the decoding procedure, we receive a binary output $\mathbf{u}$, which is the decoded codeword. The iterative procedure and the selection heuristics are summarized in Algorithm~$2$, which describes the implementation of the MP-XOR-SAT based LDPC decoder.

\begin{figure*}[t]
	\centering

  	\subfloat[]
 {\includegraphics[width=0.3\linewidth]{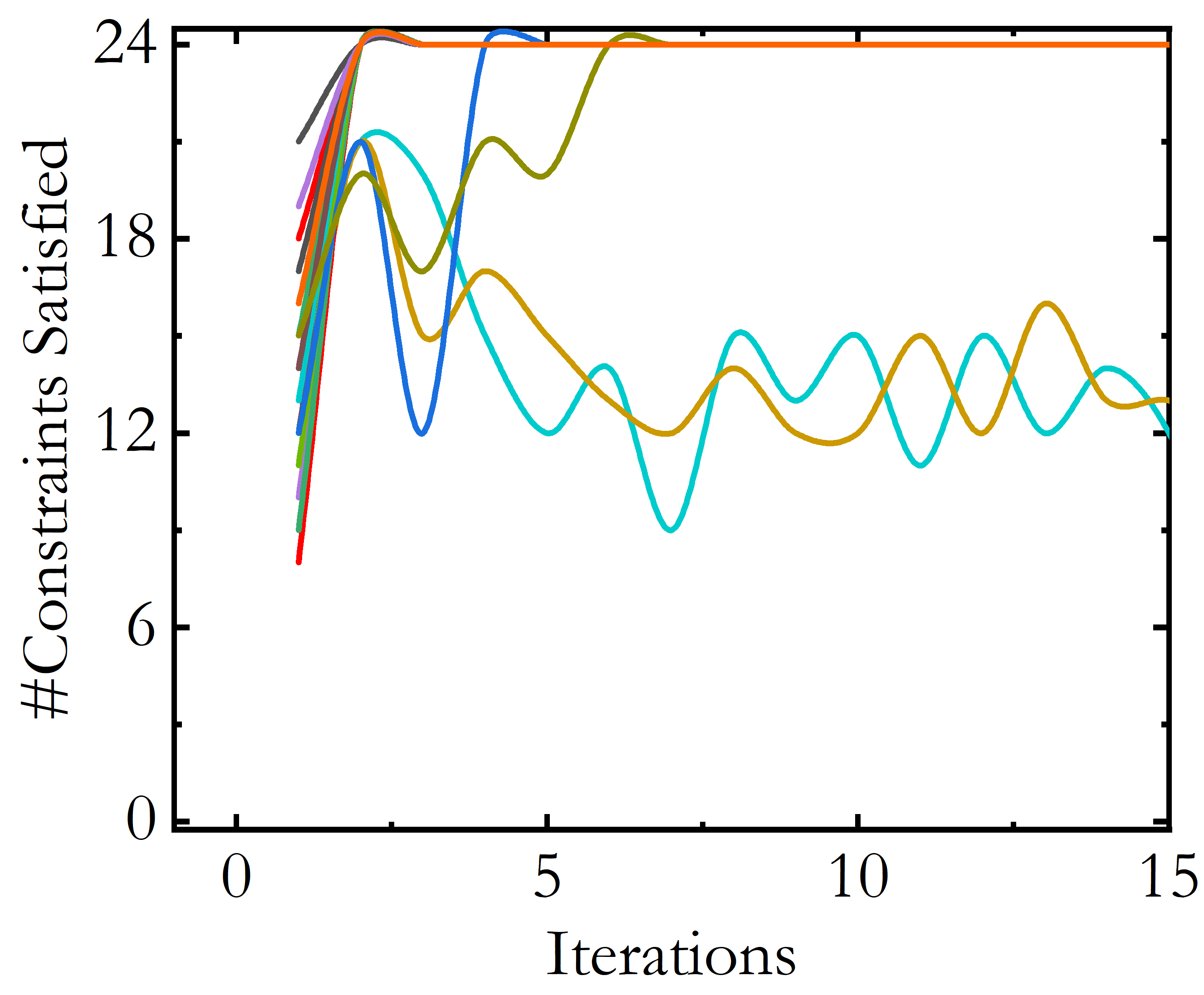}
		\label{no_reinit}}
	\subfloat[]
   {\includegraphics[width=0.29\linewidth]{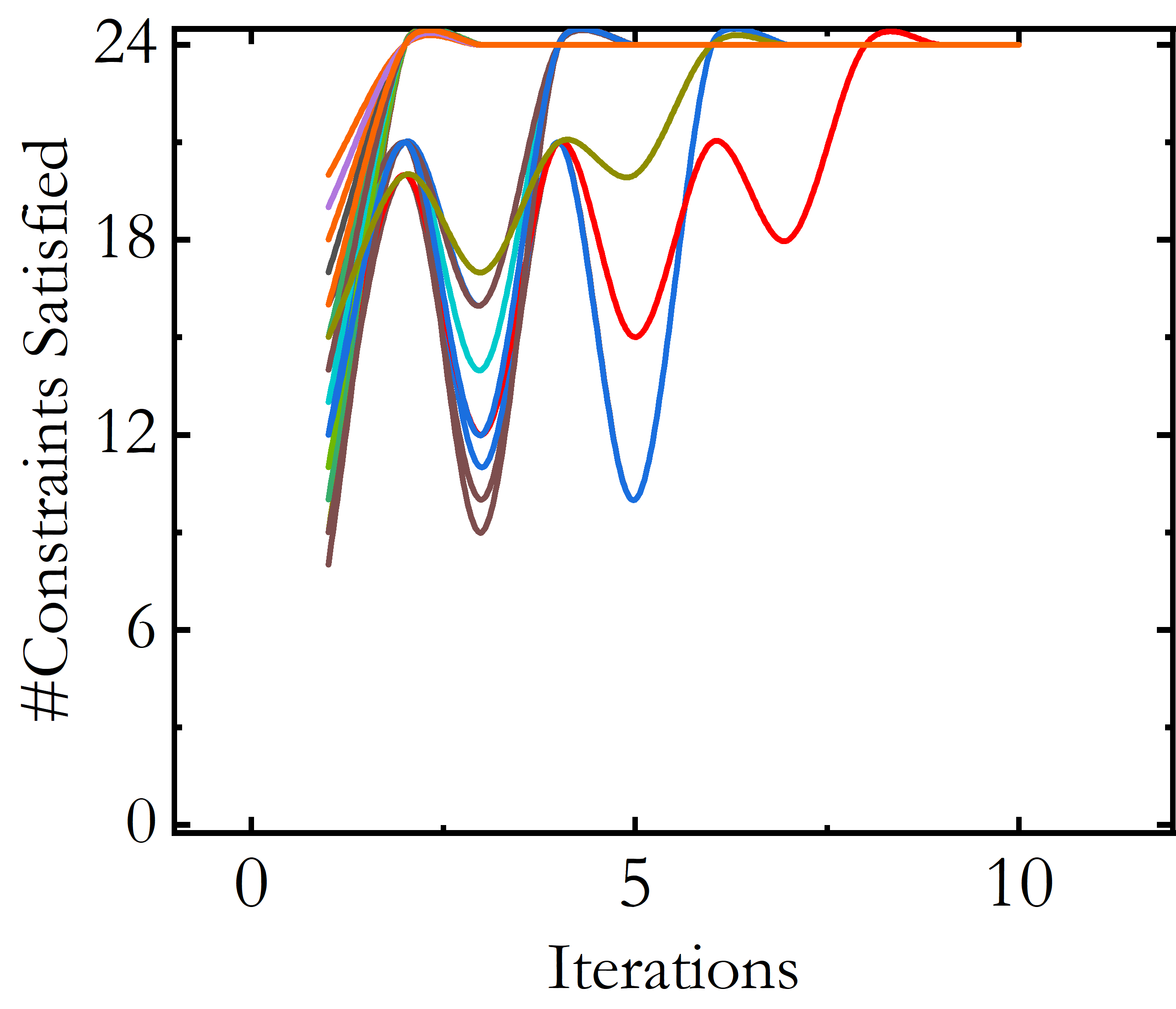}
		\label{with_re_init}}
  	\subfloat[]
   {\includegraphics[width=0.3\linewidth]{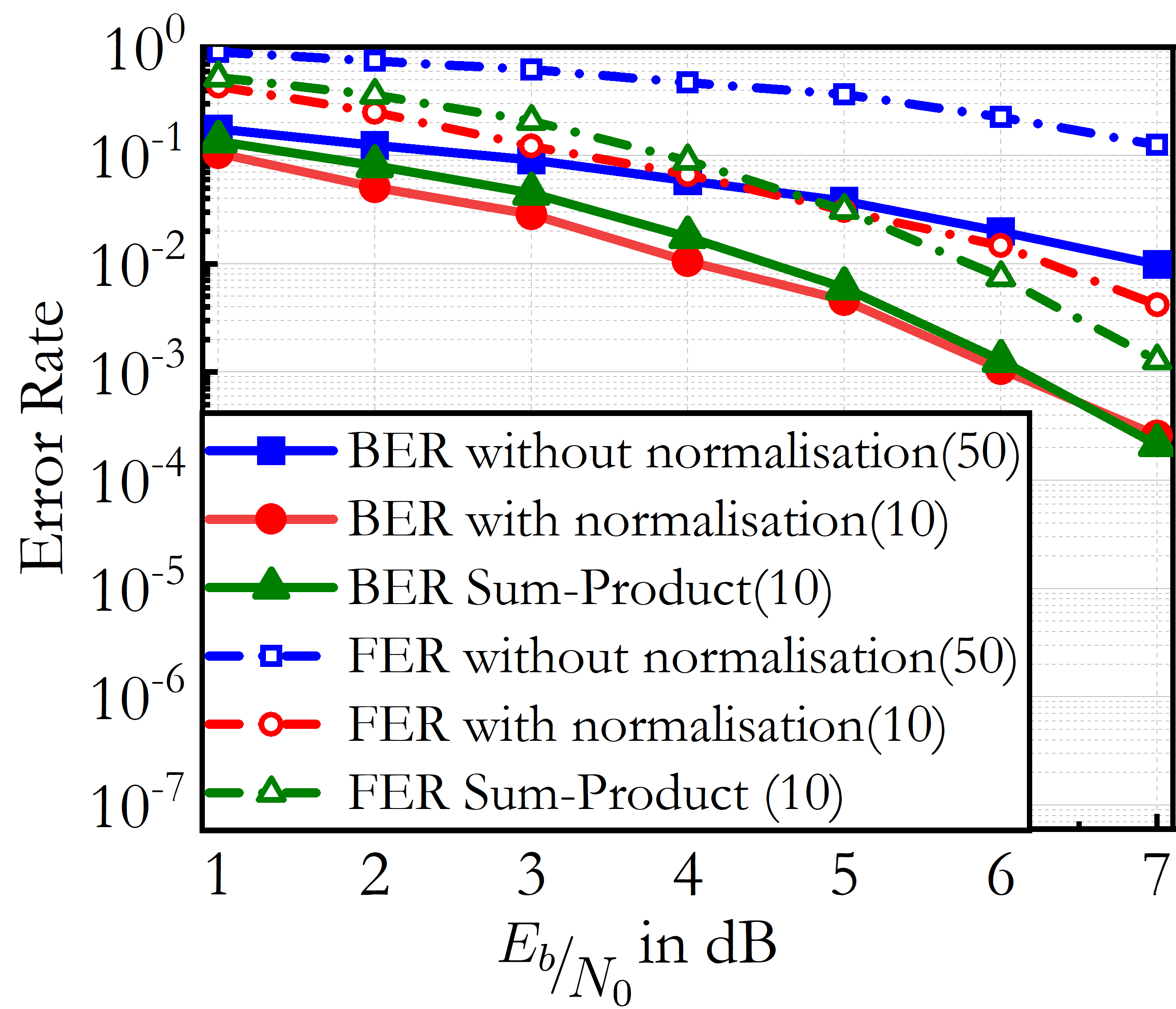}
		\label{ber_re_init}}

 	\subfloat[]
 {\includegraphics[width=0.3\linewidth]{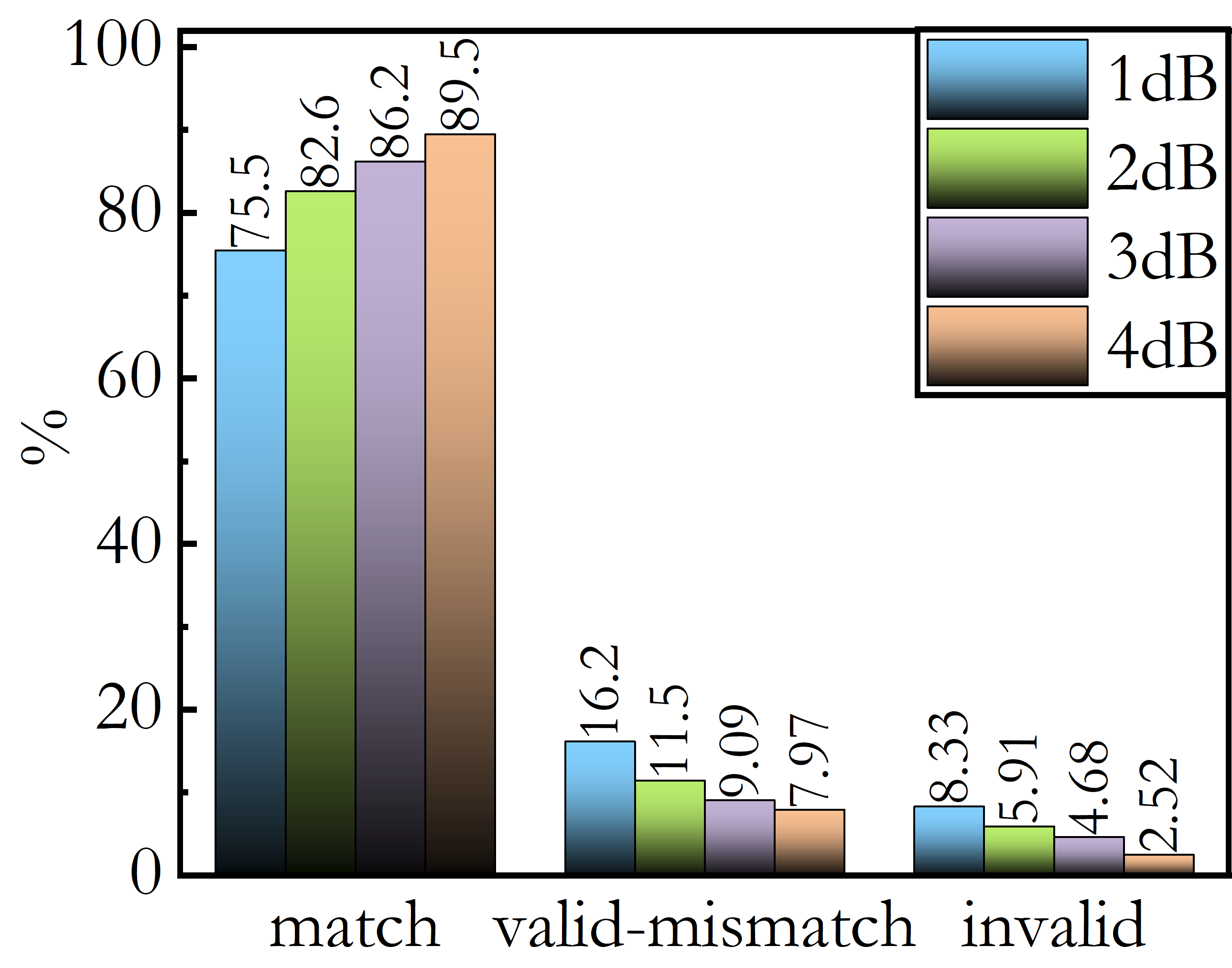}
		\label{reinit_with_50}}
	\subfloat[]
   {\includegraphics[width=0.3\linewidth]{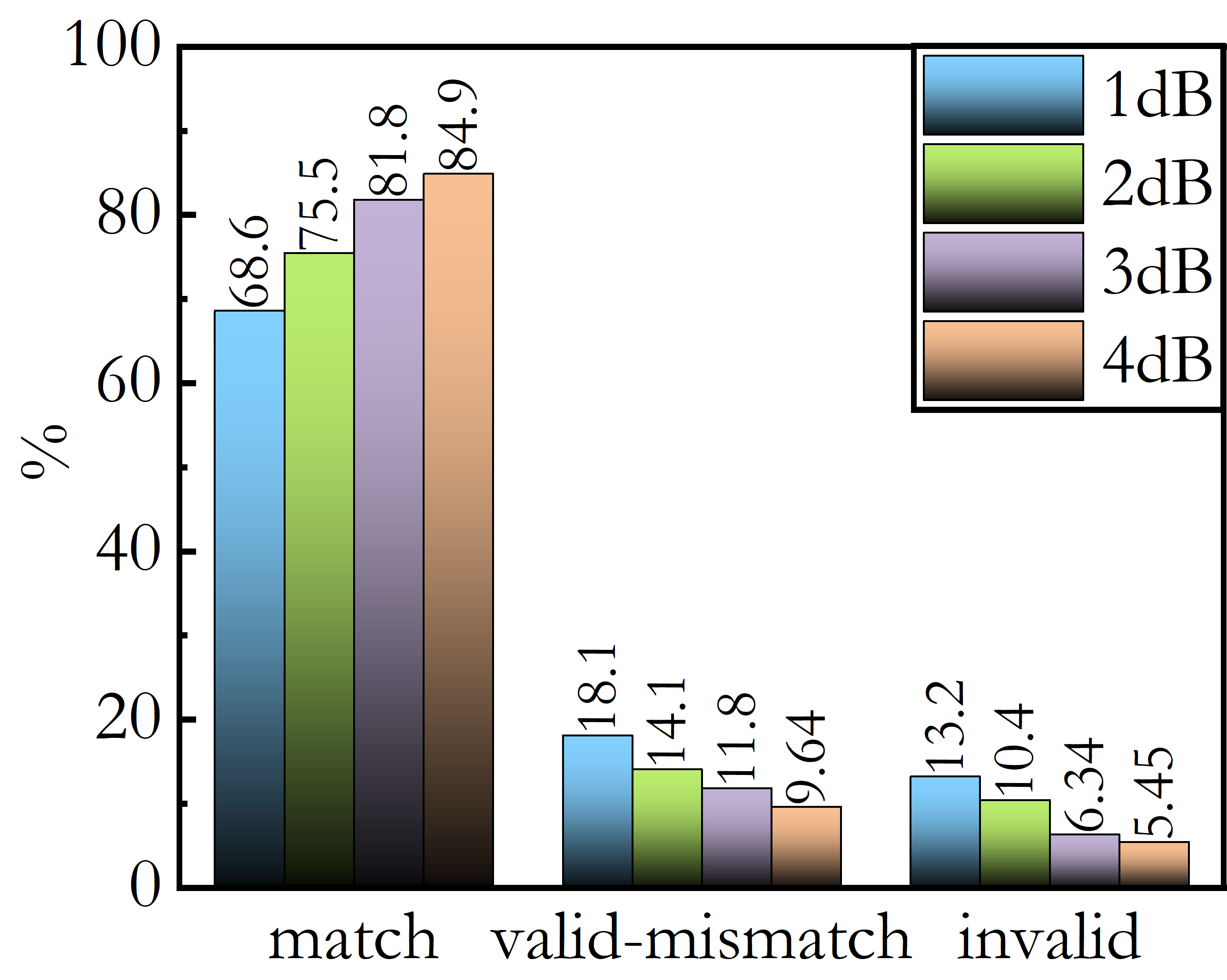}
		\label{reinit_wo_50}}
  	\subfloat[]
   {\includegraphics[width=0.3\linewidth]{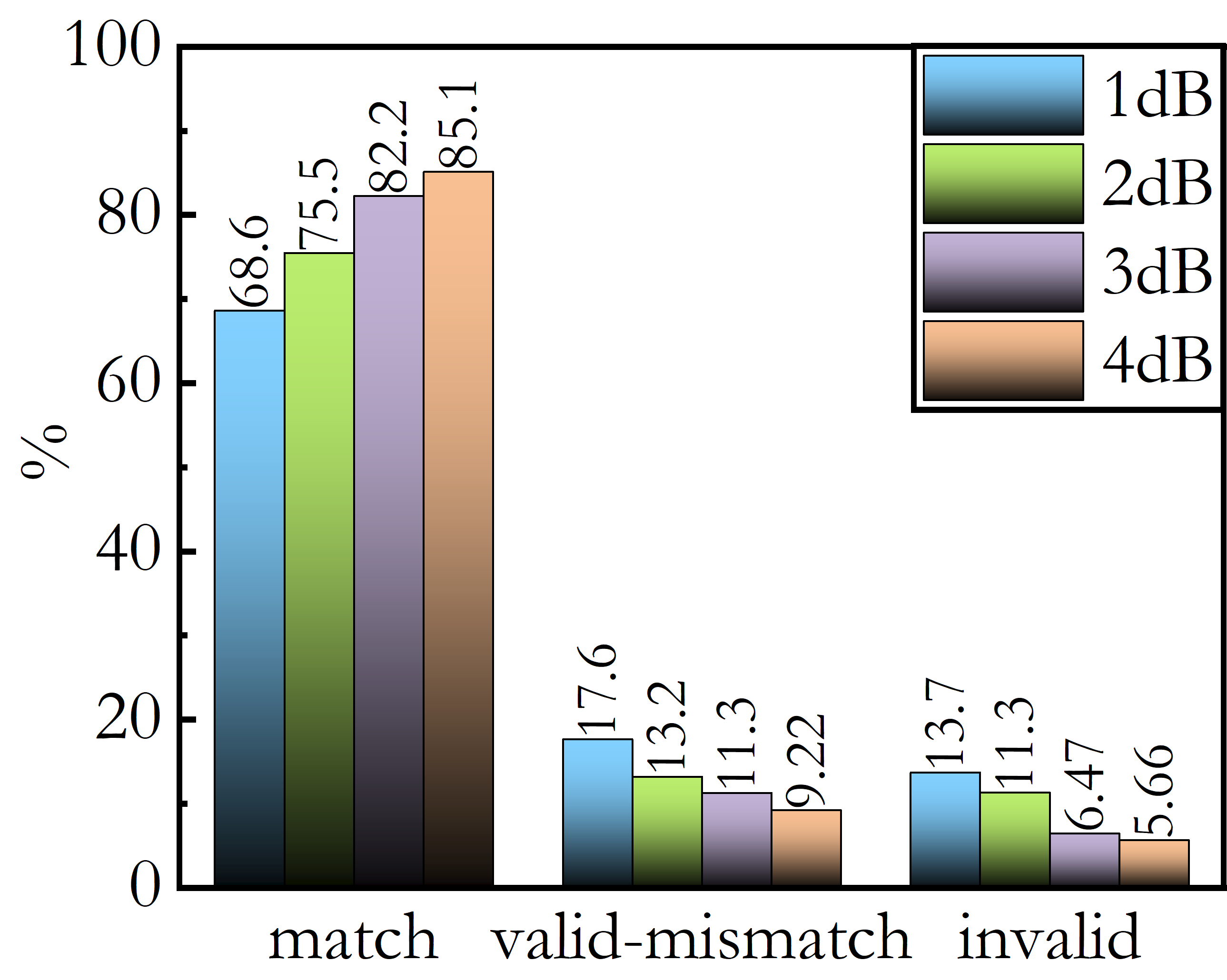}
		\label{reinit_wo_1000}}

\caption{An illustration depicting the effect of the normalization factor for a $(32,8)$ regular LDPC code for different cases with and without the normalization: \protect\subref{no_reinit} The number of constraints satisfied with respect to the iterations without normalization for $10$ random codewords at $1dB$ noise; \protect\subref{with_re_init} The number of constraints satisfied with respect to the iterations, with the normalization factor for the same $10$ noisy words in Fig.~\ref{no_reinit};~\protect\subref{ber_re_init} Bit Error Rate (BER) and Frame Error Rate (FER) with and without the normalization. Percentage of `match', `valid-mismatch' and `invalid' cases for different noise values:~\protect\subref{reinit_with_50} with the normalization factor and $I_{max}=50$;~\protect\subref{reinit_wo_50} without the normalization factor and $I_{max}=50$ iterations;~\protect\subref{reinit_wo_1000} without the normalization factor and $I_{max}=1000$. }
\label{reinit_pie}
\end{figure*}

\begin{figure*}[t]
	\centering
 	\subfloat[]
 {\includegraphics[width=0.29\linewidth]{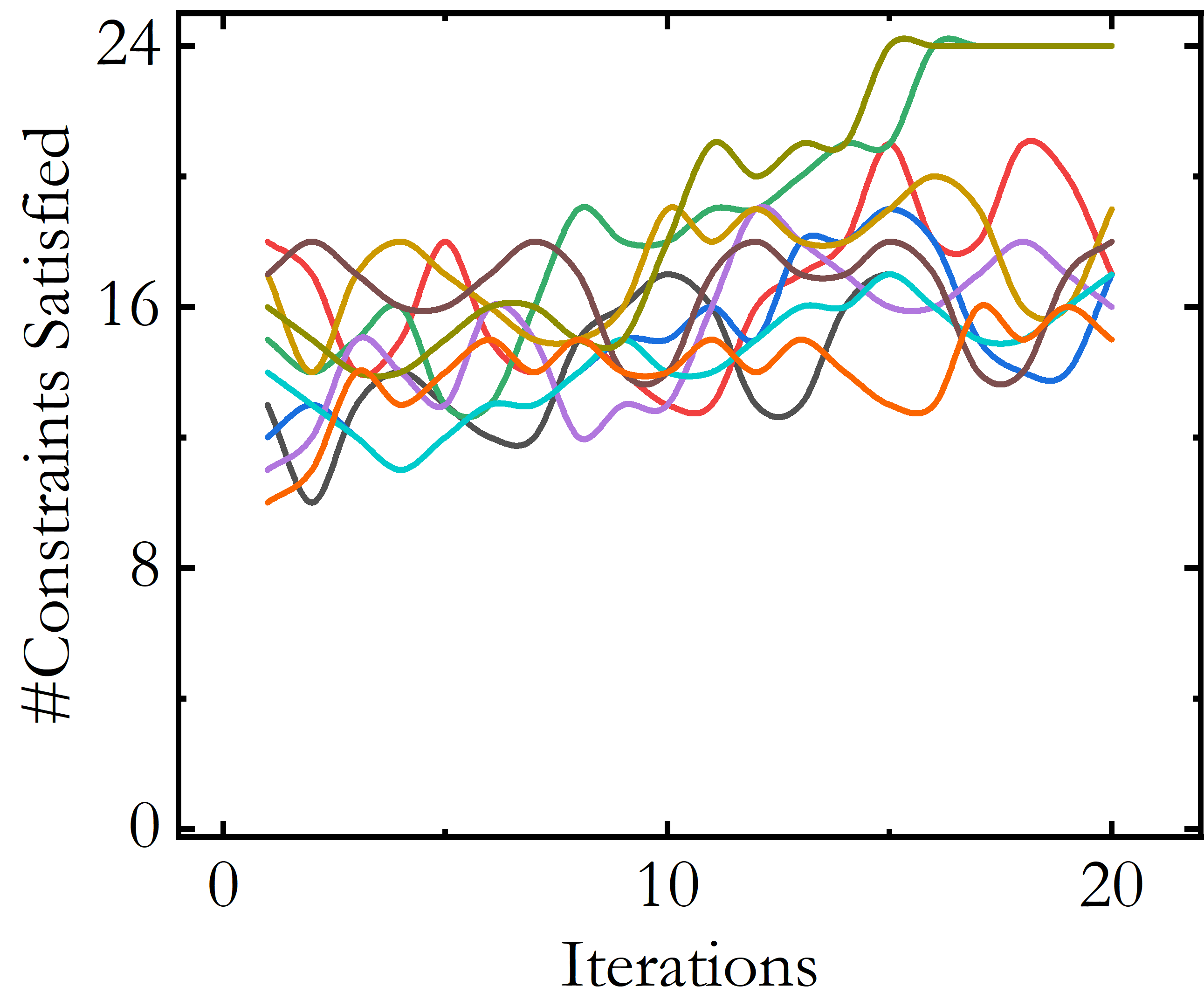}
		\label{single}}
	\subfloat[]
   {\includegraphics[width=0.3\linewidth]{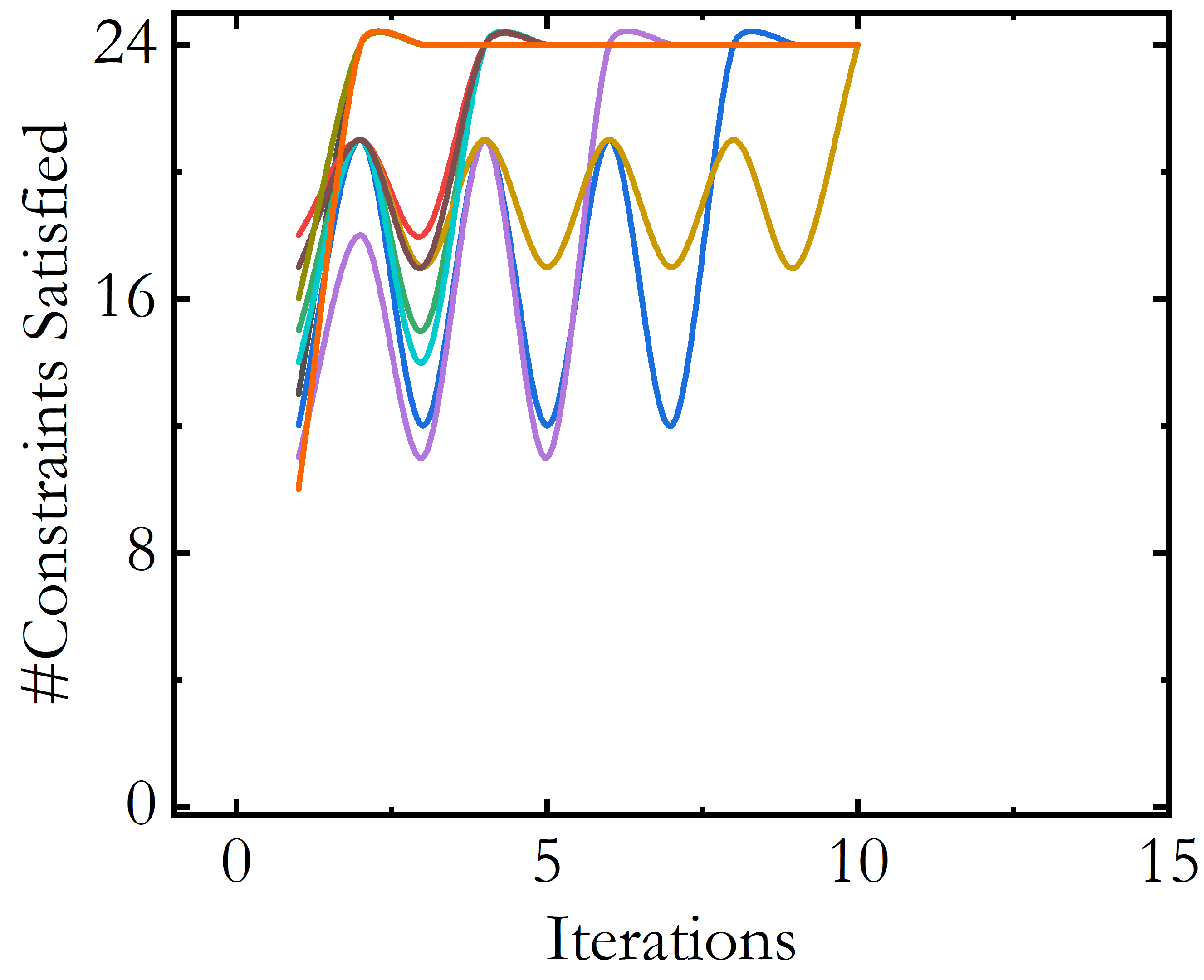}
		\label{multi}}
  	\subfloat[]
   {\includegraphics[width=0.3\linewidth]{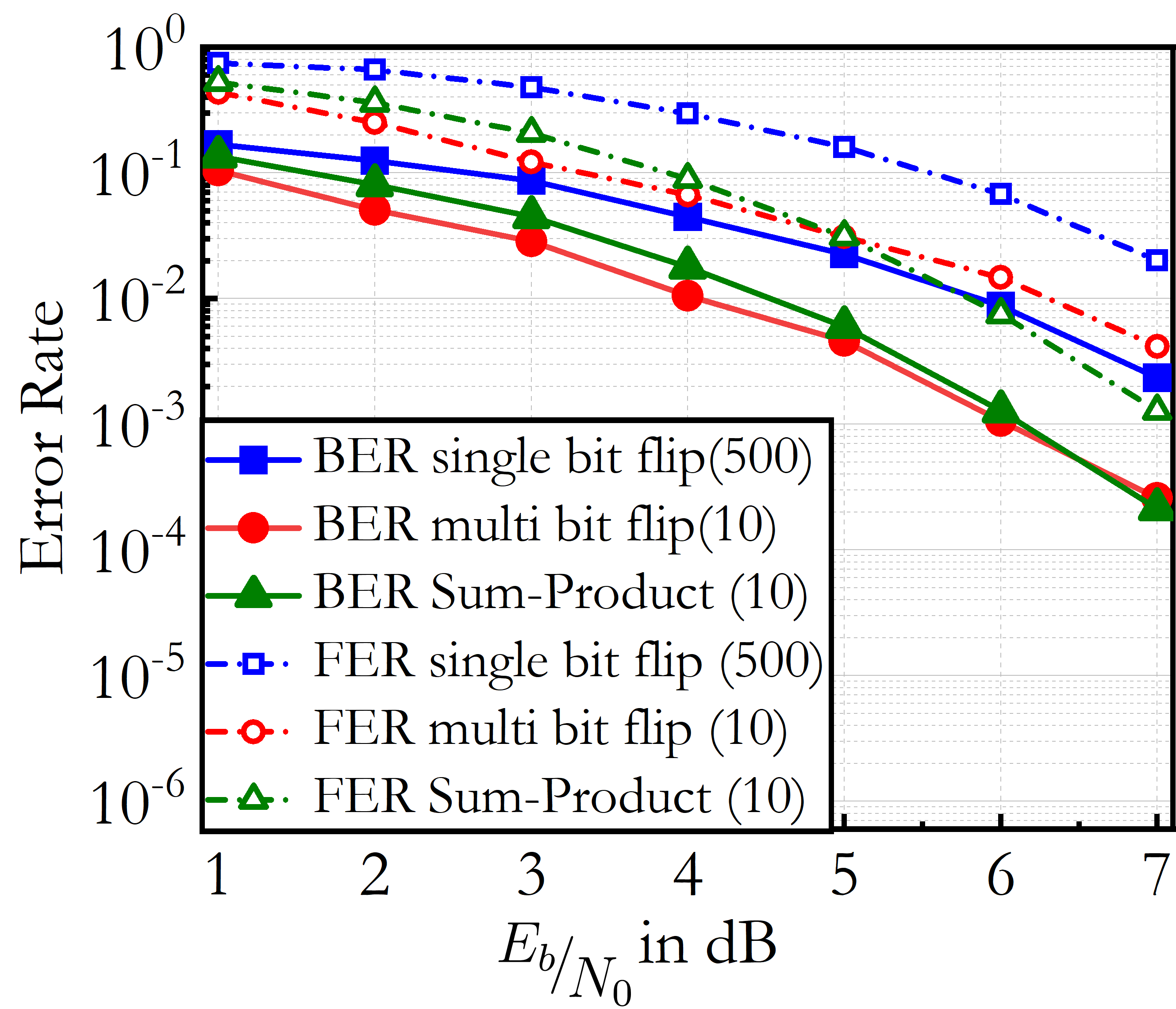}
		\label{ber_single_multi}}

\caption{An illustration showing the effect of single and multi-state bit flipping algorithm:~\protect\subref{single} A case with a single state being flipped for $10$ random codewords at $1dB$ noise;~\protect\subref{multi} A case with multiple states being flipped for the same $10$ noisy words in Fig.~\ref{single};~\protect\subref{ber_single_multi} Bit Error Rate (BER) and Frame Error Rate (FER) with single bit flip ($I_{max} = 500$) and multiple flips per iteration ($I_{max} = 10$).}
\label{bitflips}
\end{figure*}

\begin{figure*}[t]
	\centering
 	\subfloat[]
 {\includegraphics[width=0.3\linewidth]{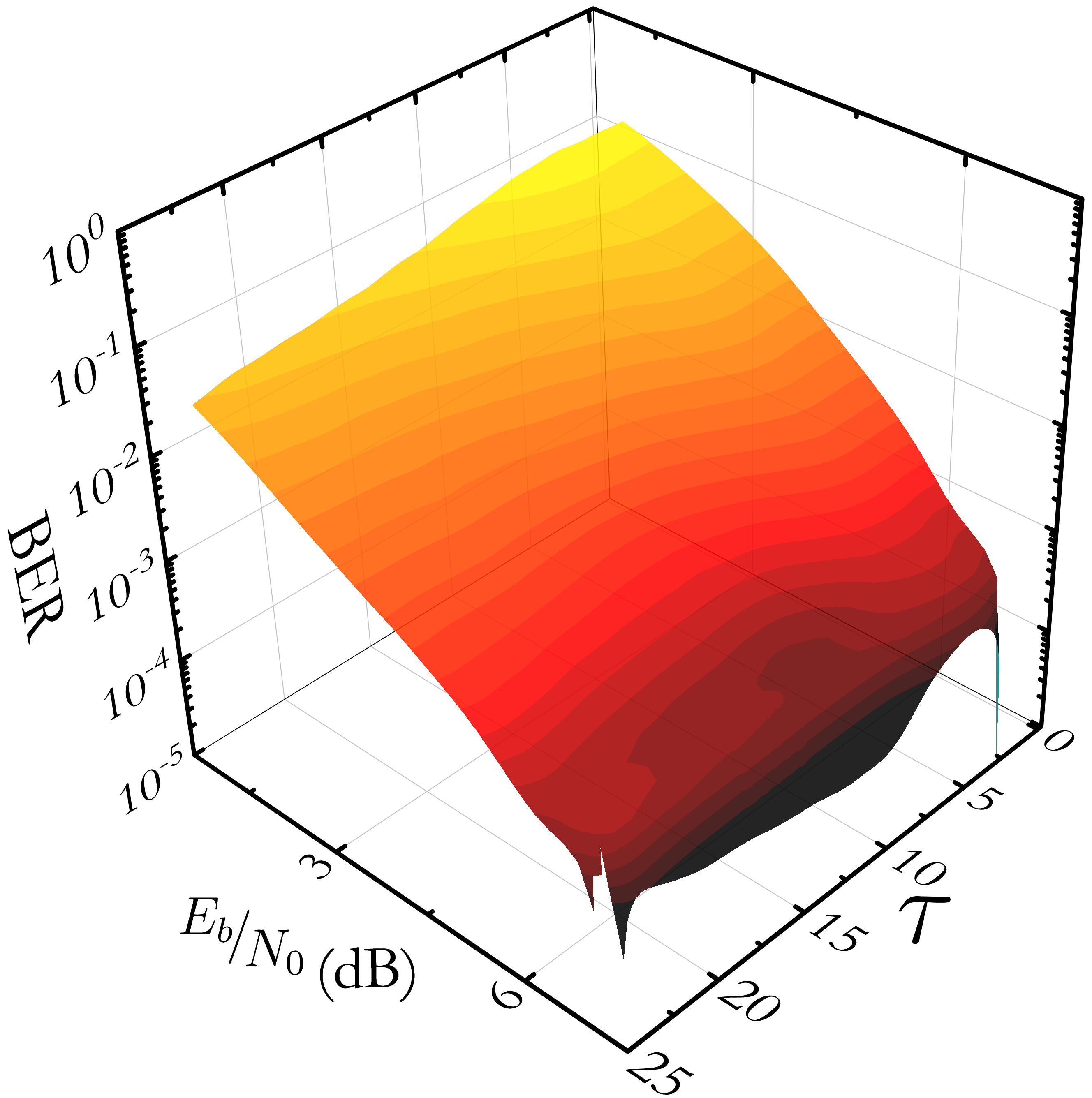}
		\label{ber_gamma}}
	\subfloat[]
   {\includegraphics[width=0.3\linewidth]{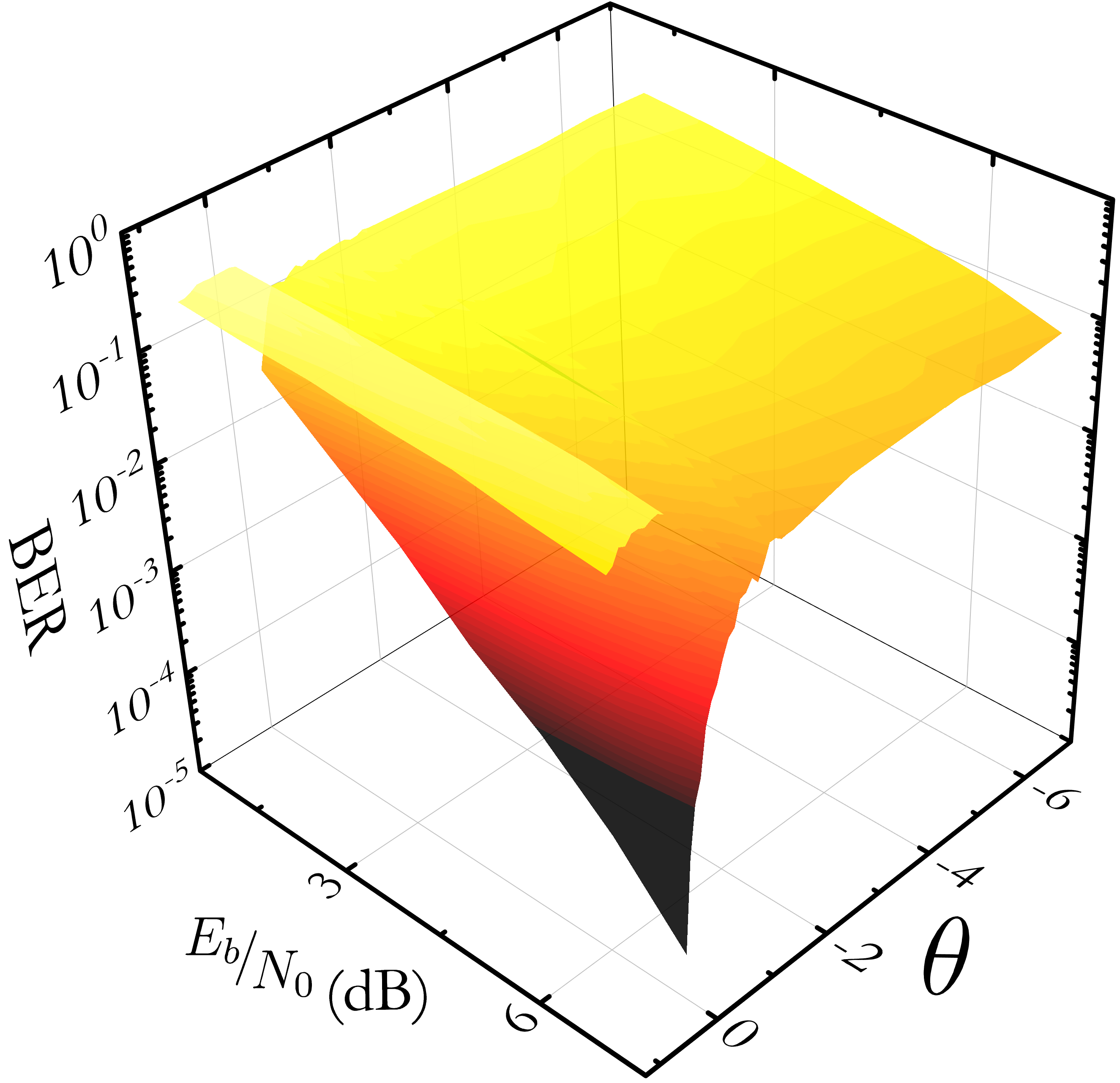}
		\label{ber_theta}}
  	\subfloat[]
   {\includegraphics[width=0.3\linewidth]{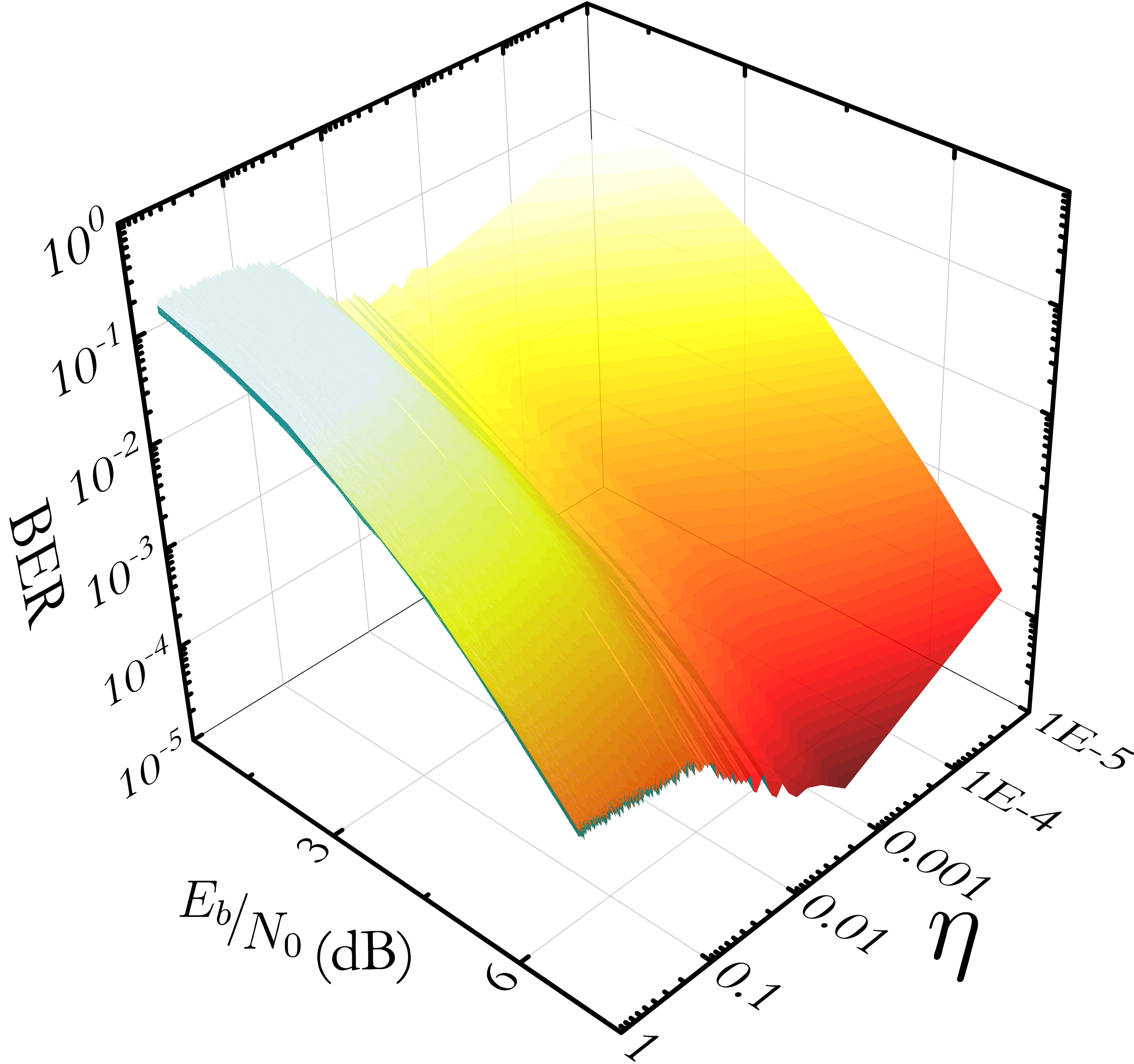}
		\label{ber_eta}}
  
  \subfloat[]
   {\includegraphics[width=0.3\linewidth]{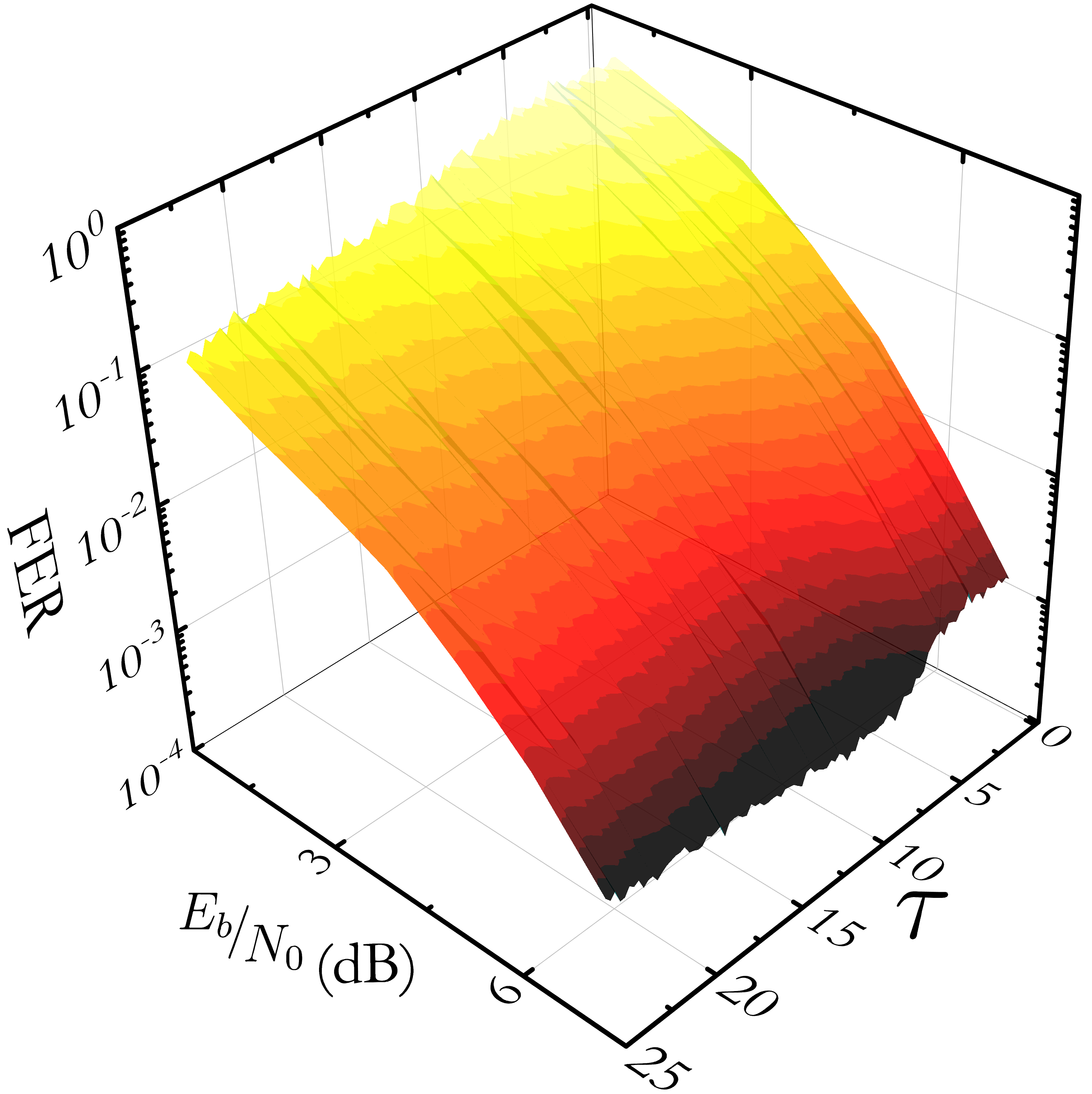}
		\label{fer_gamma}}
  \subfloat[]
   {\includegraphics[width=0.3\linewidth]{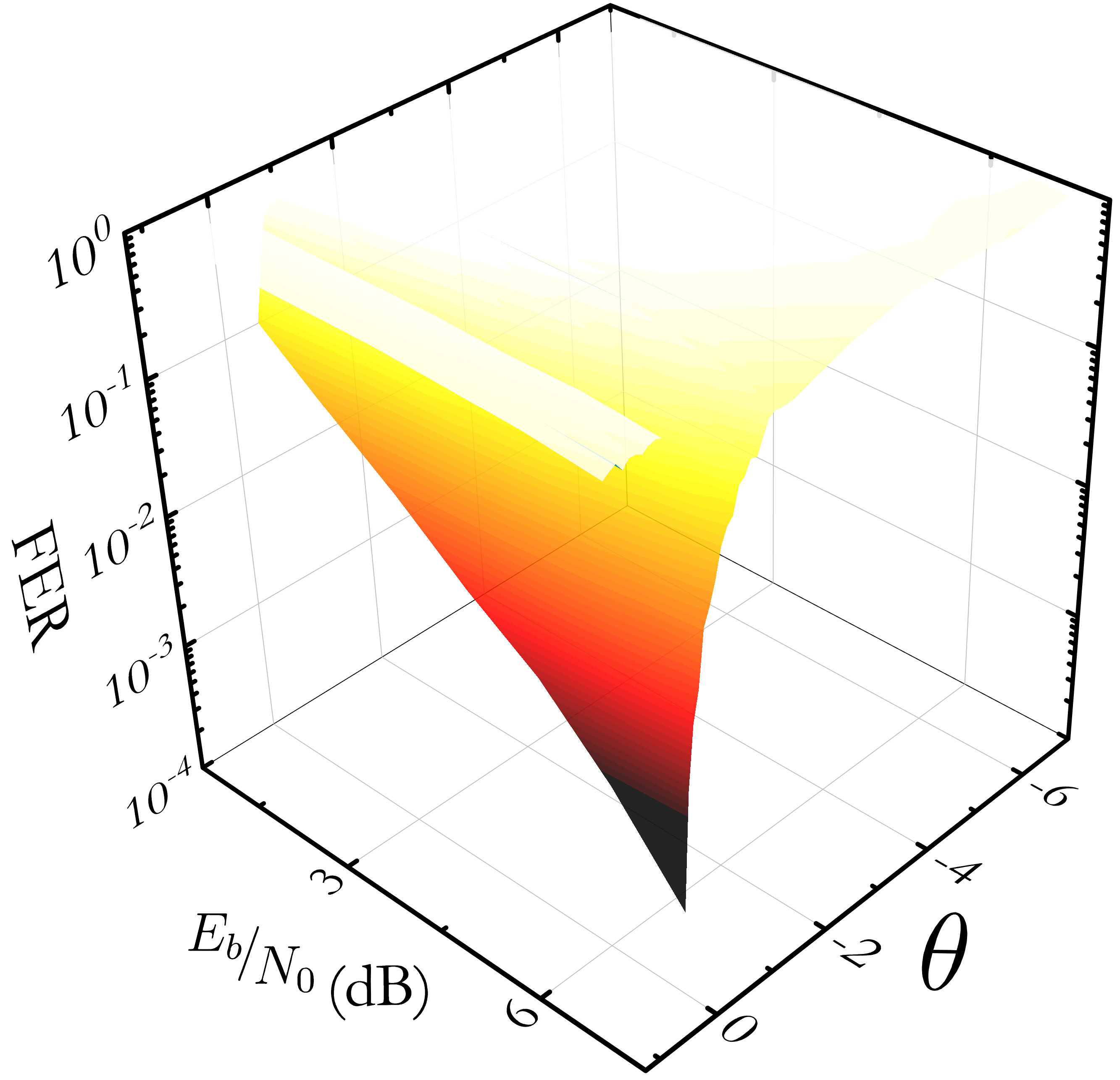}
		\label{fer_theta}}
  \subfloat[]
   {\includegraphics[width=0.3\linewidth]{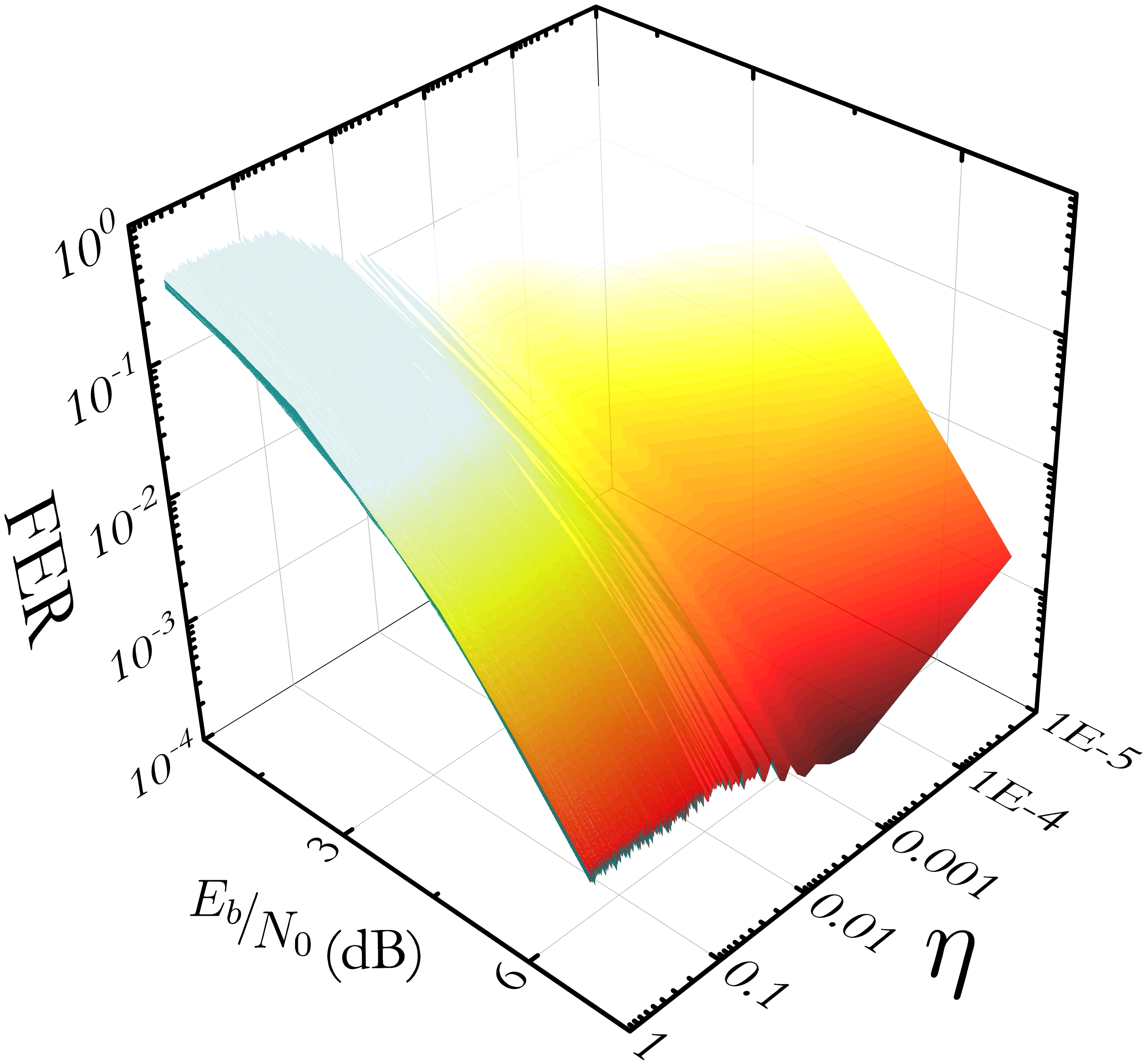}
		\label{fer_eta}}
  
\caption{An illustration showing the effect of the different hyper-parameters on the Bit Error Rate (BER) and the Frame Error Rate (FER) for different noise ($E_b/N_o$ dB) level on a (32,8) regular LDPC code. \protect\subref{ber_gamma} Represents the effect of the MP hyper-parameter $\tau$ on the BER; \protect\subref{ber_theta} Represents the optimum value of the flipping threshold $\theta$ on the BER; \protect\subref{ber_eta} Represents the optimum value of the learning rate $\eta$ on the BER; \protect\subref{fer_gamma} Represents the effect of the MP hyper-parameter $\tau$ on the FER; \protect\subref{fer_theta} Represents the optimum value of the flipping threshold $\theta$ on the FER; \protect\subref{fer_eta} Represents the optimum value of the learning rate $\eta$ on the FER.}
\label{hyperparameter}
\end{figure*}

  
  	
  



Fig.~\ref{eg_3bit} illustrates the dynamics of the MP-XOR-SAT based decoding of the $3-$bit majority code described in Fig.~\ref{concept_a}. The evolution of the bits is explained, with $d_j$ being the decision bit varying between $0$ and $1$ and $q_j$ being the soft information. The threshold is $\theta = -2.1$, and the bit flips only if the value $q_j < \theta$. The value $\tau = M$, the learning rate is $\eta = 0.5$ and the received information is $r = \left[0.1236, -1.376, 0.105\right]$. Fig.~\ref{eg_3bit_j1} describes the first bit, $j=1$. The initial value of the bit was \textit{high}, but in the $2^{nd}$ iteration, it satisfies the threshold condition and flips. This continues for the next two iterations, and the bit keeps flipping in iterations $3$ and $4$. It is only in the $5^{th}$ and last iteration when it settles. Fig.~\ref{eg_3bit_j2} describes the second bit, $j=2$. The $q$ value settles to zero, with the decision being unchanged through the iterations. In Fig.~\ref{eg_3bit_j3}, the final bit changes in the first and second iteration itself. In the third iteration, since the threshold condition is not satisfied, the bit maintains its state. It eventually settles to a \textit{low} value of the decision. Fig.~\ref{eg_3bit_constraint} shows the number of constraints that are satisfied in each constraint as the result of changing $d_j$. Before the iterative decoding procedure started, no constraints were satisfied, as referred to by iteration index $0$.  

 \section{Results}\label{results}


The first set of experiments was designed to evaluate the dynamical properties of the MP-XOR-SAT based LDPC decoder described by Algorithm~$2$ due to the normalization factor, choice of flipping scheme, and the different hyper-parameters. The second set of experiments was then designed to evaluate the Bit-Error-Rates (BER) and Frame-Error-Rates (FER) for the proposed decoder in comparison to benchmark LDPC decoding Sum-Product Algorithm with different kinds of LDPC codes.

\subsection{Effect of Normalization}
We perform an experiment to consider $10$ different codewords from a regular $(32,8)$ LDPC code with $24$ clauses. We pass these codewords through an AWGN channel and use the vector from the channel as the input to Algorithm~$2$. Fig.~\ref{no_reinit} shows the number of constraints satisfied by these ten words over the number of iterations when the normalization factor is not augmented with the cost function. Fig.~\ref{with_re_init} is presented with the normalization factor over the same ten noisy codewords. It is clearly observed that some cases may fail to converge to a solution point without normalization. We also run an experiment where we pass the codewords through an AWGN channel and send them as inputs to Algorithm~$2$ without the normalization factor with $I_{max}$ set to $50$, with the normalization factor with $I_{max} = 10$ and the sum-Product algorithm $I_{max} = 10$, where $I_{max}$ denotes the maximum allowable iterations. It can be seen from Fig.~\ref{ber_re_init} that the normalization factor improves the performance of the decoder as compared to the case without the normalization. 

We also note the number of decoded words that were valid for the given code but did not match the transmitted codeword. In reference to this, we classify three different outcomes, which are the percentages of correct decoding, i.e., the match between transmitted and decoded (denoted by `match' in the plots), the fraction that was wrongly decoded, but all the constraints were satisfied (denoted by `valid-mismatch'), and the fraction when it went down completely different trajectory (labeled `invalid'). Fig.~\ref{reinit_with_50} and Fig.~\ref{reinit_wo_50} show the percentages of the match, valid-mismatch, and invalid codewords with and without the normalization, respectively, when $I_{max} = 50$. It can be clearly observed that the percentage of matches is much higher with the normalization, and the `valid-mismatches' and `invalid' are lower. In Fig.~\ref{reinit_wo_1000}, $I_{max}$ was set to $1000$ code iterations. The results have not shown any significant improvement with respect to Fig.~\ref{reinit_wo_50}, which is shown for only $I_{max} = 50$. This shows that iterations play a negligible role as compared to the normalization factor in improving the percentage of valid mismatches.

 \subsection{Effect of Multi-bit-flips and State Update}

We observe the number of constraints satisfied for ten noisy (AWGN) codewords from a $(32,8)$ regular LDPC code, where, in one case, we flip only one bit amongst the ones that satisfy the threshold condition, while in the other case, we flip all the bits that satisfy the threshold condition in Algorithm~$2$. 

It can be observed from Fig.~\ref{single} that the trajectory of the constraints satisfied for single bit-flips with progressing iterations has very tiny hops, unlike the multi-bit flipping. The multi-bit flipping scheme hops higher, as seen in Fig.~\ref{multi}, and converges faster. Fig.~\ref{ber_single_multi} shows the influence of the bit-flipping strategy on the convergence speed. The single-bit flipping error rate waterfall curves are far away from the Sum-Product Algorithm despite having $500$ iterations. Thus, the multi-bit flipping strategy is favorable for faster decoding.

\begin{figure*}[t]
	\centering
 	\subfloat[]
 {\includegraphics[width=0.5\linewidth]{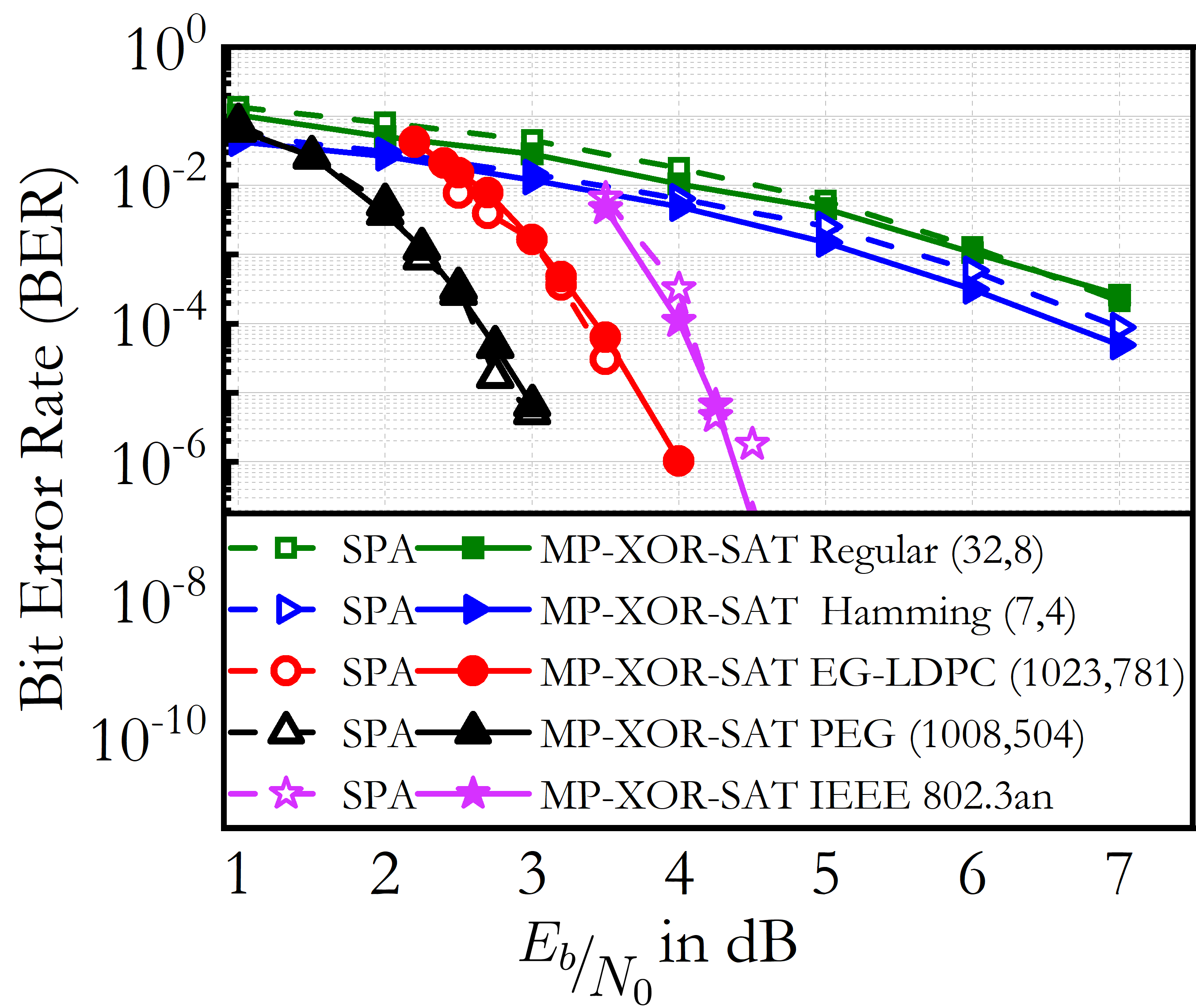}
		\label{ber_all}}
	\subfloat[]
   {\includegraphics[width=0.5\linewidth]{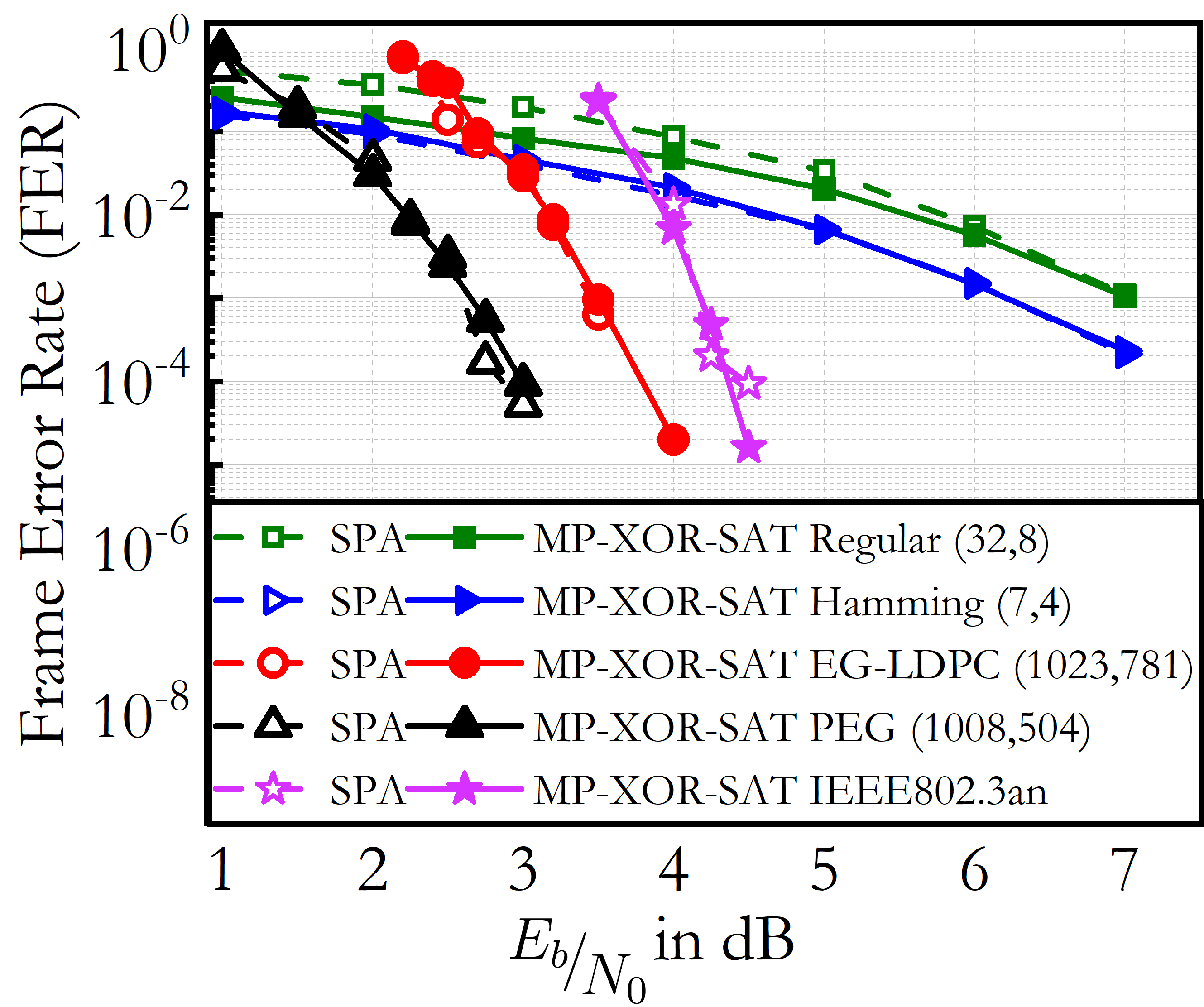}
		\label{fer_all}}

\caption{Simulation plots for an AWGN Channel for $(32,8)$ regular LDPC, $(7,4)$ Hamming, $(1023,781)$ EG-LDPC, $(1008,504)$ PEG, $(2048,1723)$ IEEE 802.3 an codes using the proposed MP-XOR-SAT algorithm; \protect\subref{ber_all} Bit Error Rate (BER) and \protect\subref{fer_all} Frame Error Rate (FER) }
\label{ER_plots}
\end{figure*}

\subsection{Effect of MP Hyper-parameters}
We compute the function's gradient using the MP gradient, which is dependent on the choice of $\tau$. We illustrate the bit and frame error rates for different values of the MP hyperparameter $\tau$ for the $E_b/N_0$ (dB) values spanning from $1$dB to $7$dB to in Fig.~\ref{ber_gamma} and Fig.~\ref{fer_gamma} for a ($32,8$) regular LDPC code. It is seen that when the $\tau$ values are close to the number of clauses, lower error rates are achieved throughout the spectrum. Thus, we chose $\tau = M$ for all our simulations.

Similar to the choice of $\tau$, the choice of the flipping threshold $\theta$ and learning parameter $\eta$ are equally important for accuracy. Fig.~\ref{ber_theta} and Fig.~\ref{fer_theta} show that the flipping threshold, when close to $0$, results in better BER and FER. If $\theta > 0$ is chosen, the results are catastrophic. Similarly, for better BER and FER, the learning rate $\eta$ must be defined between $0.01$ to $0.001$ as seen in Fig.~\ref{ber_eta} and Fig.~\ref{fer_eta}.

 \subsection{BER and FER Performance}
 Fig.~\ref{ER_plots} shows the results of the LDPC decoding using Algorithm~$2$ on an Additive White Gaussian Noise (AWGN) Channel. We present the results for Hamming ($7,4$), Regular ($32,8$)~\cite{ldpc_minggu}, EG-LDPC ($1023, 781$)~\cite{book_shu_lin}, PEG ($1008, 508$)~\cite{mackay2005encyclopedia}, and IEEE803.an~\cite{ad-gdbf} standards for Bit Error rate (BER) in Fig.~\ref{ber_all} and Frame Error rate (FER) in Fig.~\ref{fer_all}. Each code is run until $100$ erroneous codewords are listed for both the proposed algorithm and the Belief Propagation (Sum-Product Algorithm~\cite{sarah_johnson_2009}) for the same number of code iterations. It can be clearly observed from the plots that the bit and frame error plots are within $0.1$dB of the Belief Propagation algorithm, which is, by far and, to the best of our knowledge, the fastest formulation among soft decision Gradient-Descent Bit Flipping family of algorithms. 
\begin{table*}[t]
\centering
\caption{Comparison of Inversion Functions}
\scalebox{0.9}{
\begin{tabular}{|c|c|c|c|c|ccc|c|c|}
\hline
Algorithm  & Channel & Iterations & \begin{tabular}[c]{@{}c@{}}Improvement \\ wrt BP\end{tabular} & \begin{tabular}[c]{@{}c@{}}Code\\ Length (bits)\end{tabular} & \multicolumn{3}{c|}{Inversion Function Parameters}    & \begin{tabular}[c]{@{}c@{}}Flipping \\ Rule\end{tabular}          & \begin{tabular}[c]{@{}c@{}}Bit \\ Flip\end{tabular} \\ \hline
           & & & & & \multicolumn{1}{c|}{$\alpha$}  & \multicolumn{1}{c|}{$\xi(\cdot)$}        & $\lambda_j$   &  &                                                              \\ \hline
GDBF\cite{GDBF_wadayama} & AWGN & 100 & No & PEG(1008,504)  & \multicolumn{1}{c|}{1}                            & \multicolumn{1}{c|}{$\prod\limits_{j \in \mathcal{N}(i)} x_j$}     & 0                                                                                      & L1 norm                                                           &\begin{tabular}[c]{@{}c@{}}Single \\ Multi\end{tabular}                                             \\ \hline
Multi-bit\cite{multibitGDBF} & AWGN & 300 & No & PEG(1008,504)    & \multicolumn{1}{c|}{1}  & \multicolumn{1}{c|}{$\prod\limits_{j \in \mathcal{N}(i)} x_j$}     & 0                                                                                      & \begin{tabular}[c]{@{}c@{}}AWGN PDF \\  Analysis\end{tabular}                                                 & \begin{tabular}[c]{@{}c@{}}Single \\ Multi\end{tabular}                                              \\ \hline
NGDBF\cite{ngdbf} & AWGN &  300 & No & PEG(1008,504)   & \multicolumn{1}{c|}{1}      & \multicolumn{1}{c|}{$\prod\limits_{j \in \mathcal{N}(i)} x_j$}    & \begin{tabular}[c]{@{}c@{}c@{}c@{}}random noise \\ $\mu = 0$ \\ $\sigma = k^2 N_0/2$\\ $k=\left( 0,1 \right )$ \end{tabular} & \begin{tabular}[c]{@{}c@{}}Adaptive \\ Threshold\end{tabular}     & \begin{tabular}[c]{@{}c@{}}Single \\ Multi\end{tabular}     
\\ \hline


ISBF\cite{isbf} & BSC & 300 & Yes & dv4R030N1296  & \multicolumn{1}{c|}{1}  & \multicolumn{1}{c|}{\begin{tabular}[c]{@{}c@{}}$\sum s_k$ \\ $s_k = x_k \oplus y_k$\end{tabular}}  & \multicolumn{1}{c|}{$S_k$} & \multicolumn{1}{c|}{\begin{tabular}[c]{@{}c@{}}Threshold and \\ probability $p_0$\end{tabular}}  &  Multi    
\\ \hline

GDBFwM\cite{gdbf_wm} &\begin{tabular}[c]{@{}c@{}}AWGN\\ BSC\end{tabular} &300 & \begin{tabular}[c]{@{}c@{}} Yes (AWGN)\\ No (BSC) \end{tabular} & \begin{tabular}[c]{@{}c@{}c@{}} 1296(3,6) \\ 1296(4,8) \\ 2048(6,32) \end{tabular} & \multicolumn{1}{c|}{$>0$}   & \multicolumn{1}{c|}{$\prod\limits_{j \in \mathcal{N}(i)} x_j$}     & \begin{tabular}[c]{@{}c@{}c@{}c@{}}momentum \\ $\Delta x_n^{(l)}$ \\ $=x_j^{(I)} - x_j^{(I-1)}$\\ @ each iteration \end{tabular} & \begin{tabular}[c]{@{}c@{}c@{}c@{}}$E_{th}$\\ $= E_{min}+\delta$ \\ $\delta > 0 $\\$p<1$\end{tabular}     & \begin{tabular}[c]{@{}c@{}}Single \\ Multi\end{tabular}     
\\ \hline

AD-GDBF~\cite{ad-gdbf}  & BSC & 300-1000 & Yes & \begin{tabular}[c]{@{}c@{}c@{}} 1296(iRISC) \\ 1296(5GNR) \\ 2048(IEEE 802.3) \end{tabular} & \multicolumn{1}{c|}{$>0$}  & \multicolumn{1}{c|}{$\prod\limits_{j \in \mathcal{N}(i)} x_j$} & momentum & \begin{tabular}[c]{@{}c@{}}$E_{th} = E_{min} +\delta$ \\ $\delta >0, p<1$\end{tabular} & Multi  
\\ \hline 

This Work  & AWGN & 30-100 & Yes & \begin{tabular}[c]{@{}c@{}c@{}} PEG (1008,504) \\ 1023(EG-LDPC) \\ 2048(IEEE 802.3) \end{tabular} & \multicolumn{1}{c|}{$\{-1,+1\}$}  & \multicolumn{1}{c|}{$\sum\limits_{j \in \mathcal{N}(i)} q_j$} & $0$ & \begin{tabular}[c]{@{}c@{}}Experimental \\ Threshold\end{tabular} & Multi  \\ \hline
\end{tabular}}
\label{comparison_table}
\end{table*}


\begin{table*}[ht]
\centering
\caption{Complexity Analysis per Iteration}
\begin{tabular}{|c|c|c|c|c|c|}
\hline
          & Additions & Multiplications & Comparators  & XOR & Others \\ \hline
GDBF~\cite{GDBF_wadayama} & $N\gamma$         & $M\left(\rho - 1\right)+N$     & $1$  & -      &   -     \\ \hline

NGDBF~\cite{ngdbf} & $3N$  & $N$ (binary)  & $-$   & - & \begin{tabular}[c]{@{}c@{}c@{}} Counters $(2N)$\\  Random Number \\ Generator $(1)$\end{tabular} \\ \hline

ISBF~\cite{isbf} &  $>N$  & $-$   &  $>N$    & $N+M(\rho-1)$ &  - \\ \hline

GDBFwM~\cite{gdbf_wm} &  $\left(\gamma+2\right)N+1$   & $2N$    &  $4N-1$  & $-$ & \begin{tabular}[c]{@{}c@{}} Random Number \\ Generator \end{tabular} \\ \hline

AD-GDBF~\cite{ad-gdbf}   & \begin{tabular}[c]{@{}c@{}} $\left (2+ \beta \right)N$ \\ $0 \leq \beta \leq 2$ \end{tabular}         & $-$             &  $\left (2N-1\right)$   & $N+M(\rho-1)$ &\\ \hline
This Work & $(M\gamma + N\rho +1)$  & $-$    & $N$    &  $M(\rho-1)$ & \begin{tabular}[c]{@{}c@{}} $M-$input\\MP function$(2)$\end{tabular} \\ \hline
\end{tabular}
\label{Complexity Analysis}
\end{table*}

\section{Discussion}\label{discussion}

\subsection{Algorithm Performance}
As has been discussed earlier, the normalization factor aims to maintain close proximity with the initial assumption, similar to the Maximum Likelihood decoding algorithm. Without the normalization, the bit flips tend to dissatisfy more constraints, causing the dynamics to take a different trajectory. Normalization ensures fewer valid mismatches in the decoding as they are more difficult to recover than invalid words. Literature suggests the usage of variety decoders (adaptive with varying hyper-parameters for a batch of code iterations, or with belief propagation decoders, etc.) in cascade to improve the decoding performance~\cite{ad-gdbf,gdbf_2023}. The same can be employed to treat the invalid output to improve the decoding performance further.

The investigation of single-state bit flipping and multi-state bit flipping strategies has been a subject of exploration in numerous literature sources, with the majority suggesting the expedited convergence of multi-bit flipping strategy. Our findings align consistently with the conclusions drawn in prior studies. 

To the best of our knowledge, the proposed MP-XOR-SAT solver is the fastest bit-flipping formulation in terms of code iteration and error rates compared to the Sum-Product algorithm. Our survey summarizes the number of code iterations taken by different GDBF decoders in Table~\ref{comparison_table} along with an indicator if the decoder performs better than Belief Propagation (BP). AD-GDBF~\cite{ad-gdbf} does better than BP in $300-1000$ iterations for the Binary Symmetric Channel, but its performance has not been listed on the AWGN channel. 

\subsection{Objective Function}

As has been mentioned earlier, the nature of the proposed algorithm based on XOR-SAT solvers has a close resemblance with Gradient Descent Bit Flipping algorithms. Gradient Descent Bit Flipping (GDBF) algorithms have been under exploration for almost two decades, and quite naturally, there are multiple variations while trying to achieve better performance with reduced complexities. In general, the entire plethora of these algorithms can be generalized. For the energy associated with the $j^{th}$ variable, we write a generalized equation~\eqref{generalized_obj}, where $\alpha$ is a scaling weight and $\xi(\cdot)$ is a function as in~\cite{GDBF_wadayama, multibitGDBF, ngdbf, gdbf_wm, ad-gdbf} or form of the received signal $x_j$~\cite{isbf}. The term $\lambda_j$ can be a random variable~\cite{ngdbf} or function (momentum in~\cite{gdbf_wm}).  

\begin{equation}
    E(x_j) = \alpha  x_jr_j + \sum_{i \in \mathcal{M}\left(j\right)} \xi(\cdot) + \lambda_j
    \label{generalized_obj}
\end{equation}

When the parameters $\alpha$, $\xi(\cdot)$, and $\lambda_j$ take various values, we get different variations of GDBF. For the basic GDBF\cite{GDBF_wadayama}, $\alpha=1$, $\xi(\cdot)=\prod_{j\in N\left(i\right)} x_j$ and $\lambda_j=0$. Here, the notation $\mathcal{N}(i) \triangleq \{j\in \left[1, N\right]: h_{ij}=1\}$ denotes the parity-check neighborhood, i.e., the set of variable nodes associated with the $i^{th}$ check node. 
Table~\ref{comparison_table} summarizes the literature on Gradient Descent Bit Flipping algorithms. It can be clearly observed that works dealing with the AWGN channel output have to solve the product of the output channel, while the ones using BSC take the summation of the hard decision output, clearly indicating the complexity increase in the case of AWGN decoders. 

The scaling factor $\alpha$ is chosen differently in different works and cannot be generalized. The term $\lambda_j$ has a large diversity ranging from random samples to momentum functions. The energy function of this work, however, cannot be directly analyzed by the above form. It can be visualized to be the maximization of the difference between the logarithm of energy functions of satisfying clauses and the dissatisfied ones, as seen from the form of $\mathcal{H}_{log}$. 
The functions of both these complimentary clauses are of the form mentioned in equation~\eqref{generalized_obj}. 
The appropriate parameters are mentioned in Table~\ref{comparison_table}.

\subsection{Complexity Analysis}
There are many different Gradient Descent Bit Flipping algorithms as described in Table~\ref{comparison_table}. However, we limit our analysis to recent Gradient Descent Bit Flipping algorithms and those that have considered soft decision decoding with the AWGN channel in the past and summarize it in Table~\ref{Complexity Analysis}. It is clear from Table~\ref{comparison_table} that the algorithms dealing with AWGN channels are far more complex due to their multiplicative nature than the ones dealing with BSC channels, as seen for~\cite{isbf, ad-gdbf}.

The proposed objective function is modeled to function on the logarithmic values of the channel output, thus eliminating the use of complex computations like multiplication and division, unlike the others. However, there is an added overhead of the number of additions, which is much simpler than multiplication. There is also a drastic decrease in the number of comparators as compared to the others. Given that the proposed formulation computes the variable bits using summation, the XOR gates are only required to identify which check clauses satisfy and which check clauses dissatisfy the constraints. Thus, $M(\rho-1)$ $2-$input XOR gates are required. 

The other overhead is of the MP function, which, however, is a two-transistor block for a single input~\cite{synth}. The output is analog and, if designed in the current domain, can eliminate the need for any circuit to implement the additions. This analysis shows that implementing this in the analog or mixed signal domain can yield tremendous advantages in power and area of the circuit; however, this will be addressed in future work.

\section{Conclusion}\label{conclusion}

In summary, this paper introduces a novel soft decision Margin Propagation-based XOR-SAT solver, which can be used for LDPC decoding. The algorithm, sharing similarities with Gradient Descent Bit Flipping methods, is thoroughly examined in terms of its objective function, which maximizes the clauses satisfying the constraints. After extensive analysis, it can be observed that most of the earlier decoding algorithms were one or the other form of the original GDBF algorithm proposed by Wadayama et al. in~\cite{GDBF_wadayama}. However, the objective function of the proposed work has a different form and leads to faster convergence of the code words, i.e., $10\times$ lesser code iterations than the GDBF decoders in literature. Results from decoding performance simulations on an AWGN channel demonstrate the algorithm's competitiveness with the Sum-Product Algorithm, with Bit and Frame Error rates within $0.1$dB while taking the same number of code iterations. The complexity analysis highlights the algorithm's efficiency, especially in terms of reduced multiplication operations and comparators, suggesting its potential for implementation in the analog or mixed-signal domain. In conclusion, the XOR-SAT solver-based decoding algorithm presents a promising solution for efficient LDPC decoding. Future work will focus on practical implementations in the analog domain and further optimizations to enhance its feasibility for chip design.

\section*{Acknowledgment}
The authors would like to acknowledge the United States India Educational Foundation (USIEF) and Institute of International Education (IIE) for sponsoring the in-person research at Washington University in St. Louis via the Fulbright Nehru Doctoral Fellowship. This work is also supported by the Prime Ministers' Research Fellowship.

\bibliographystyle{IEEEtran}
\bibliography{ref}

\end{document}